\documentclass[12pt]{article}
\RequirePackage[OT1]{fontenc}
\usepackage{amsmath, amsthm, amssymb}
\usepackage{bm, dsfont}
\usepackage{natbib}
\usepackage[margin=1in]{geometry}
\usepackage{color}
\usepackage{graphicx}
\usepackage{hyperref}

\newcommand{\NN}{\mathbb{N}}

\newcommand{\RR}{\mathbb{R}}
\newcommand{\ZZ}{\mathbb{Z}}

\DeclareMathOperator{\EE}{\mathbb{E}}
\DeclareMathOperator{\PP}{\mathbb{P}}
\DeclareMathOperator{\Var}{Var}

\newcommand{\calB}{\mathcal{B}}
\newcommand{\calL}{\mathcal{L}}

\newcommand{\one}{\mathds{1}}

\DeclareMathOperator*{\argmax}{arg\,max}

\newtheorem{thm}{Theorem}[section]

\newtheorem{prop}[thm]{Proposition}

\theoremstyle{definition}
\newtheorem{defi}[thm]{Definition}
\newtheorem{rem}[thm]{Remark}

\newtheorem*{condition}{Condition}

\begin{document}

\thispagestyle{empty}
\vskip 5mm
\begin{center} 
{\Large{\bf{Patterns in Spatio-Temporal Extremes}}}
\end{center}

\vskip 5mm

\begin{center}
\large
Marco Oesting$^1$ and Rapha\"el Huser$^2$ 
\end{center}

\footnotetext[1]{
\baselineskip=10pt Stuttgart Center for Simulation Science (SC SimTech) \& Institute for Stochastics and Applications, University of Stuttgart,
70569 Stuttgart, Germany, E-mail: marco.oesting@mathematik.uni-stuttgart.de}
\footnotetext[2]{
\baselineskip=10pt Statistics Program, Computer, Electrical and Mathematical Sciences and Engineering (CEMSE) Division, King Abdullah University of Science and Technology (KAUST), Thuwal 23955-6900, Saudi Arabia. E-mail: raphael.huser@kaust.edu.sa}

\vskip 4mm
\centerline{\today}
\vskip 6mm

\begin{center}
{\large{\bf Abstract}}
\end{center}
In environmental science applications, extreme events frequently exhibit a complex spatio-temporal structure, which is difficult to describe flexibly and estimate in a computationally efficient way using state-of-art parametric extreme-value models. In this paper, we propose a computationally-cheap non-parametric approach to investigate the probability distribution of temporal clusters of spatial extremes, and study within-cluster patterns with respect to various characteristics. These include risk functionals describing the overall event magnitude, spatial risk measures such as the size of the affected area, and measures representing the location of the extreme event. Under the framework of functional regular variation, we verify the existence of the corresponding limit distributions as the considered events become increasingly extreme. Furthermore, we develop non-parametric estimators for the limiting expressions of interest and show their asymptotic normality under appropriate mixing conditions. Uncertainty is assessed using a multiplier block bootstrap. The finite-sample behavior of our estimators and the bootstrap scheme is demonstrated in a spatio-temporal simulated example. Our methodology is then applied to study the spatio-temporal dependence structure of high-dimensional sea surface temperature data for the southern Red Sea. Our analysis reveals new insights into the temporal persistence, and the complex hydrodynamic patterns of extreme sea temperature events in this region.

\par\vfill\noindent

\pagenumbering{arabic}

\newpage

\allowdisplaybreaks
\clearpage

\section{Introduction}

Many environmental extreme events, such as heavy precipitation \citep{castrocamilo-huser-20}, heat waves \citep{zhong-etal-22}, or storms \citep{oesting-etal-17}, have a complex spatio-temporal structure that has a direct influence on their statistical properties. With appropriate statistical methods, this structure can be revealed by the use of spatio-temporal data, which have become available in increasingly higher resolutions due to ongoing technical progress. At the same time, the relevant events of interest are by definition rare, so observations usually contain only a few (if any) individual extreme events, which makes traditional empirical approaches inefficient (if useful at all).
An alternative more reliable approach for tail extrapolation is to use specialized parametric models backed up by extreme-value theory. Such models include max-stable processes \citep{dehaan-84,davison-etal-12,davison-huser-15,davison-etal-19}, $r$-Pareto processes \citep{ferreira-dehaan-14,dombry-ribatet-15,thibaud-opitz-15,defondeville-davison-18}, or more recent processes that possess higher flexibility at sub-asymptotic levels \citep{wadsworth-tawn-12,huser-wadsworth-19,wadsworth-tawn-22}; see \citet{huser-wadsworth-22} for a recent review of these parametric modeling approaches. However, one faces two major challenges with such parametric approaches: it is both difficult to develop (i) models that are relatively parsimonious and yet flexible enough to capture complex tail dependence structures, and (ii) fast inference methods to fit these models efficiently with large datasets.

With regard to challenge (i), the extreme-value statistics community has so far primarily focused on developing either purely temporal or purely spatial models. Only in recent years, truly spatio-temporal models have been emerging \citep[see, e.g.,][]{davis-etal-13,huser-davison-14,defondeville-davison-20,Simpson.etal:2022}, but they are often restricted to be stationary, spatially isotropic, space-time separable, or subject to irrealistic dependence assumptions \citep[see, e.g.,][]{huser-wadsworth-22,hazra-etal-21}. In most cases, the resulting models mainly aim at representing the marginal distributions and the pairwise dependencies with maximum accuracy, often disregarding higher-order dependence interactions. Certain crucial characteristic features of the events caused by the underlying physical processes, such as a storm growing, moving over an area, and weakening, are indeed very difficult to describe flexibly using parametric models. 

With regard to challenge (ii), such spatio-temporal parametric extreme-value models are often computationally intensive to fit in high dimensions, especially when they are based max-stable processes \citep{padoan-etal-10,castruccio-etal-16,huser-etal-19}. Recent progress has been made to fit $r$-Pareto processes efficiently using a gradient-scoring approach \citep{defondeville-davison-18}, and to make fast inference for the conditional extremes model using \texttt{R-INLA} \citep{Simpson.etal:2022}. Amortized inference approaches based on neural Bayes estimators have also been recently advocated to fit max-stable and other extreme-value processes efficiently \citep{sainsburydale-etal-22}. Nevertheless, developing truly flexible spatio-temporal models that can reliably capture a variety of joint tail events of interest and within-cluster behaviors, and that can be fitted in truly high dimensions (e.g., of the order of $5000\times 11000$ dependent spatio-temporal locations as in our data application in Section~\ref{sec:application}) remains currently inaccessible. 

In this paper, instead of relying on parametric models, we propose an alternative and computationally-cheap non-parametric approach that relies on mild standard assumptions about the regularity of the joint tail (important for reliable tail extrapolation), and which allows us to efficiently answer a wide variety of key scientific questions, such as
\begin{itemize}
    \item[(Q1)] How long does an extreme event last?
    \item[(Q2)] Does the magnitude of the event increase/decrease within its course?
    \item[(Q3)] Does the spatial area affected by the event shrink/grow?
    \item[(Q4)] Does the ``center'' of the extreme event move in a certain way?
\end{itemize}
While these questions are quite basic, the answers may be not straightforward, even through appealing visual plots. 
\begin{figure}[t!]
        \centering
	\includegraphics[width=0.29\textwidth]{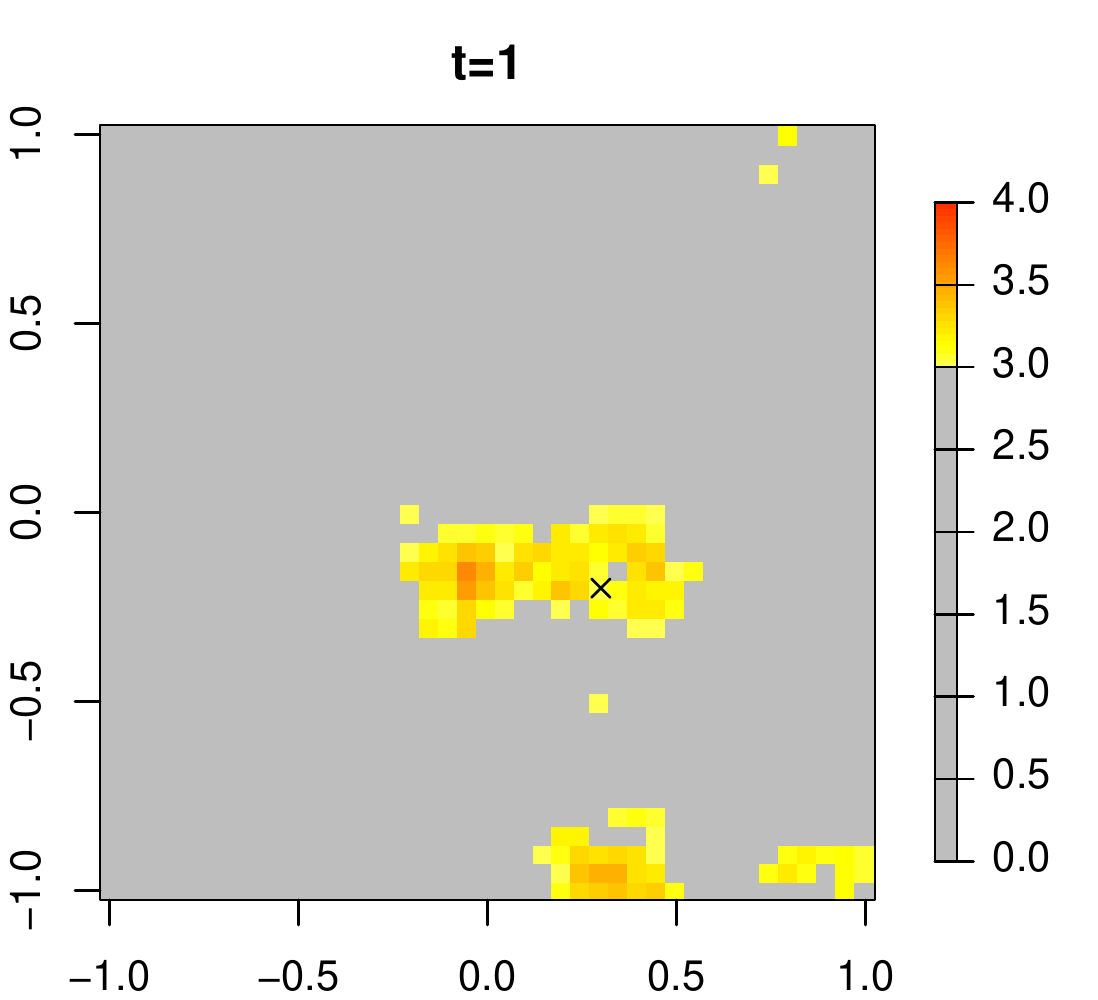}
 	\includegraphics[width=0.29\textwidth]{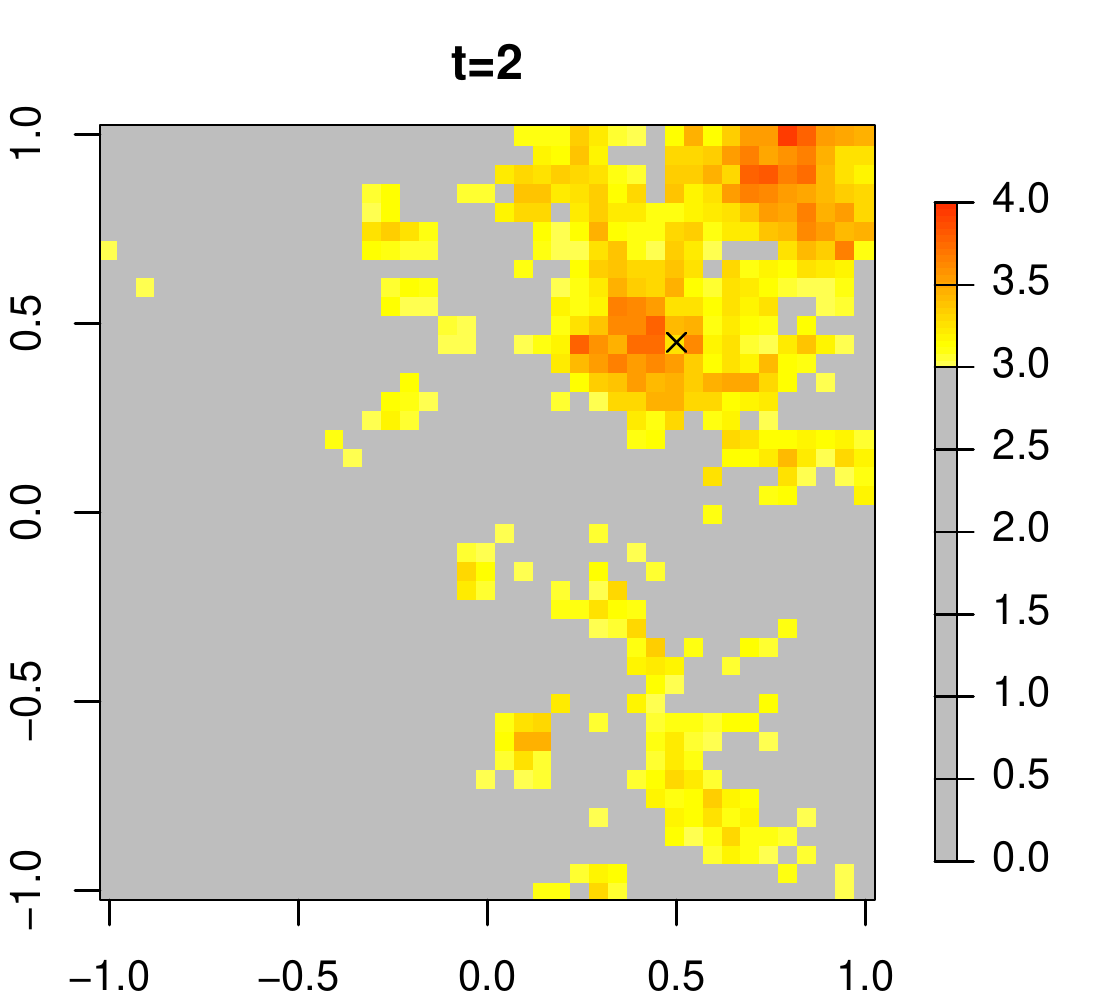} 
  	\includegraphics[width=0.29\textwidth]{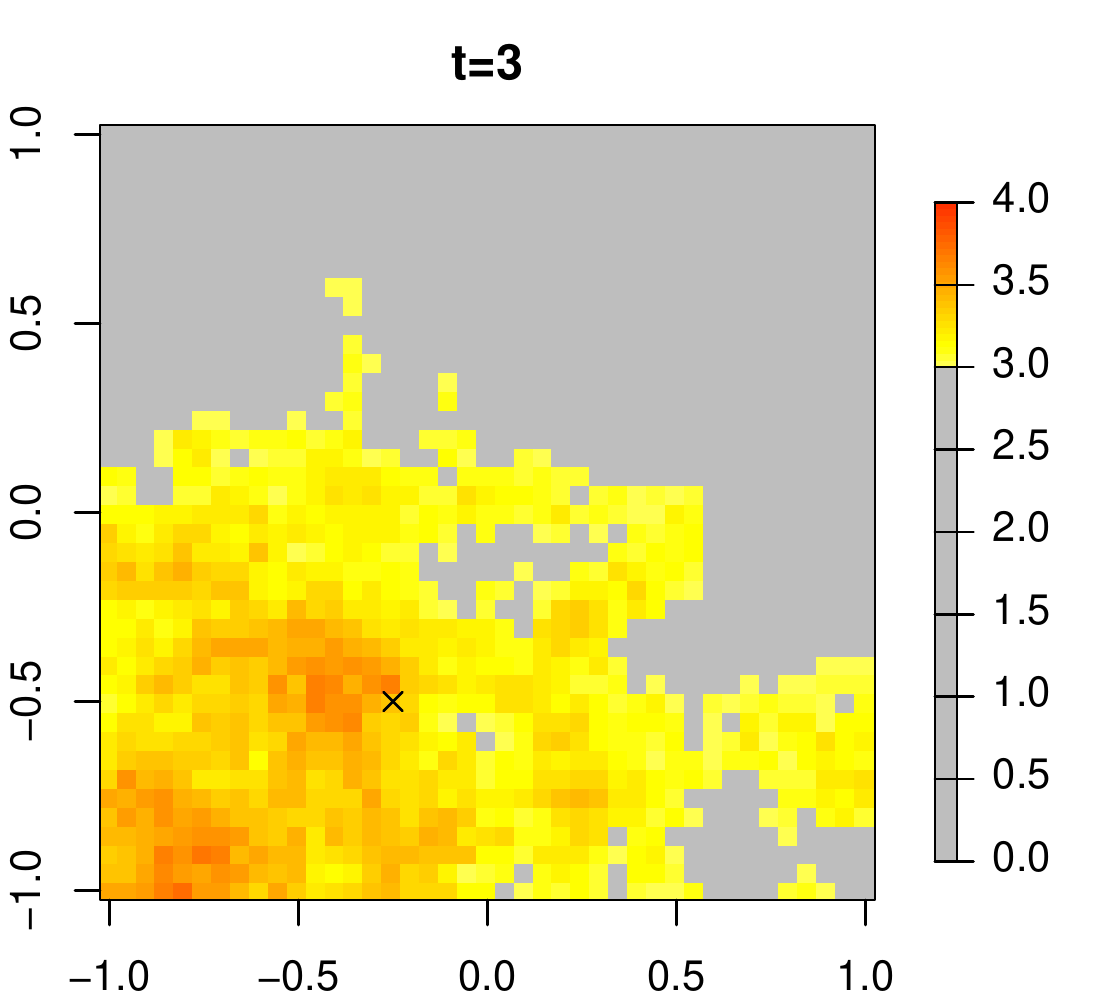} 
	\caption{An extreme event as a realization of a spatio-temporal random field at three instances of time ($t=1$, $t=2$ and $t=3$; from left to right). The exceedances of the critical threshold $u$ (here $u=3$) are displayed in colors. The black crosses mark the componentwise median of the area affected by the extreme event.} \label{fig:illustration}
\end{figure}
To illustrate this, Figure~\ref{fig:illustration} displays exceedances of a critical high threshold in three consecutive time points, based on some simulated data. Thus, one might say that this is an extreme event of duration $3$ or more, which provides a partial answer to (Q1). Furthermore, 
it can be seen that the size of the area affected by the extreme event is progressively growing in the displayed period (Q3), while, as the black crosses indicate, the ``center'' of this event moves first from the center to the northeast and further in the opposite direction to the southwest (Q4). The answer to question (Q2) requires a closer look at the actual values of the magnitude, e.g., the spatial maximum, indicated by the darkest red color in the figure, which increases from $t=1$ to $t=2$ and decreases thereafter.

More careful probabilistic answers to any of the questions raised above requires a more precise definition of extreme events. Here, we employ the idea of clusters of extremes, for which various definitions already exist in the context of univariate time series \citep[see e.g.,][]{ferro-segers-03,markovich-14}, and we extend them to the functional time series setting. As the above discussion also demonstrates, the zig-zag behavior of certain characteristics of extremes are of interest to address questions such as (Q2), (Q3) and (Q4). We shall describe such a zig-zag behavior by using ordinal patterns, a popular concept in time series analysis which has found applications in various areas \citep[see][for instance]{bandt-pompe-02, keller-etal-07, sinn-etal-13}. Here, we generalize this tool to describe certain features of spatio-temporal extreme events. Our proposed methodology extends the recent approach of \cite{oesting-schnurr-19} from univariate time series to process-valued time series.

As hinted above, we need assumptions on the space-time process that allow us to reliably extrapolate beyond the range of the observed data. Our main assumption, which is quite typical in extreme-value statistics, is that the spatio-temporal process of interest, $X$, is regularly varying as a time series. As we do not only consider the process itself, but also functionals applied to the time series, we need to introduce a refined and more flexible notion of regular variation that can be applied in a more general setting. This notion is introduced in Section~\ref{sec:regvar}. In Section~\ref{sec:risk-fun-regvar}, this concept is then used to define extremes and clusters of extremes in a functional setting and to deduce limit expressions for the cluster size distribution. The within-cluster behavior, in particular with respect to risk functionals, spatial risk measures, and location measures, is further analyzed in Section~\ref{sec:inside-cluster}. In Section~\ref{sec:estim}, we then introduce specialized ratio estimators for the corresponding limit distributions, and show that they are asymptotically normal under appropriate mixing and anti-clustering conditions. In Section~\ref{sec:simu}, the finite-sample behavior of our proposed ratio estimators is illustrated in a spatio-temporal simulated example. The estimators are then used in Section~\ref{sec:application} to analyze the complex extremal behavior of sea surface temperature in the southern Red Sea. We finally conclude with some discussion and perspective on future research in Section~\ref{sec:conclusion}.

\section{Regular variation of functional time series} \label{sec:regvar}

In the following, we consider a nonnegative stochastic process $\{X(\bm s,t)\}_{\bm s \in S, t \in \ZZ}$ on a spatio-temporal domain $S \times \ZZ$ where the spatial domain $S \subset \RR^d$ is compact and the temporal domain $\ZZ$ is discrete. We assume that, for each time $t \in \ZZ$, the process $\{X(\bm s,t)\}_{\bm s \in S}$
possesses continuous sample paths, denoted by $\{X(\bm s,t)\}_{\bm s \in S} \in C_+(S)$, i.e., the time series is $\{\bm X_t\}_{t \in \ZZ}$ is $C_+(S)$-valued. For simplicity, we will assume that the time series is stationary, i.e., the space-time process $X$ is stationary in time (but not necessarily in space).

Our proposed extreme-value methodology relies on the more general concept of regular variation for functional time series. Here, we follow \citet{hult-lindskog-06} and \citet{dombry-hashorva-soulier-18} in considering regular variation in a rather broad setting for stationary time series on a metric space. More precisely, we shall study stationary time series $\{\bm X_t\}_{t \in \ZZ}$ that take values in some complete separable metric space $E$, but for simplicity, we will restrict ourselves to the following cases: 
\begin{itemize}
 \item $E = [0,\infty)$, i.e., we have a time series of random variables,
 \item $E = [0,\infty)^d$, i.e., we have a time series of random vectors,
 \item $E = C_+(S)$, i.e., we have a time series of nonnegative sample-continuous stochastic processes on some spatial domain such as mentioned above,
\end{itemize}
equipped with
the metric $d(\bm x, \bm y) = \|\bm x - \bm y\|_\infty$ for all $\bm x, \bm y \in E$.
As these spaces satisfy all the conditions listed in \citet{dombry-hashorva-soulier-18}, regular variation of an $E$-valued time series can be defined in the following way.

\begin{defi} \label{def:regvar}
  An $E$-valued stationary time series $\{\bm X_t\}_{t \in \ZZ}$ is \emph{regularly
  varying}, if, for any finite set $T \subset \ZZ$, the random vector 
  $(\bm X_t)_{t \in T}$ is regularly varying on $E^{|T|}$, i.e., if there
  exists a nondecreasing sequence $\{a_n\}_{n \in \NN}$
  tending to $\infty$ and a non-zero Radon measure $\mu_T$ on $E^T \setminus \{\bm 0\}$,
  the \emph{exponent measure}, such that
  $$ n \PP( (a_n^{-1} \bm X_t)_{t \in T} \in A)
     \stackrel{n \to \infty}{\longrightarrow} \mu_T(A) $$
  for any measurable set $A \subset E^T \setminus \{\bm 0\}$ bounded
  away from $\bm 0$ such that $\mu_T(\partial A) = 0$.	
\end{defi}

It can be shown that the sequence $\{a_n\}_{n \in \NN}$ is necessarily regularly varying with some index $1/\alpha > 0$, i.e., $a_{\lceil nt\rceil}/a_n \to t^{1/\alpha}$ for all $t>0$ as $n \to \infty$, and that the measure $\mu_T$ is positively homogeneous of order $-\alpha$ where $\alpha$ is the same for all finite $T \subset \ZZ$.
Thus, $\{\bm X_t\}_{t \in \ZZ}$ is also called regularly varying
with index $\alpha$. Note that also the sequence $\{a_n\}_{n \in \NN}$ can be chosen independently of the index set $T \subset \ZZ$, for instance, by the relation
$$ n \PP(a_n^{-1} \|\bm X_0\|_\infty > 1) \stackrel{n \to \infty}{\longrightarrow} 1. $$

Here, we repeat the characterization of regularly varying time series via the so-called tail process and spectral tail process, which are essential concepts needed to derive the asymptotic properties of our estimators defined below. For $E=\RR^d$, the result has been stated in \citet{basrak-segers-09}, Theorem 2.1, Theorem 3.1 and Corollary 3.2 and, in a very general setting, in \citet{dombry-hashorva-soulier-18}, Lemma 3.5.

\begin{thm} \label{thm:char-regvar}
  Let $\{\bm X_t\}_{t \in \ZZ}$ be an $E$-valued stationary time series. Then,
  the following statements are equivalent (where $\calL(\cdot)$ denotes ``the probability law of''):
  \begin{enumerate}
  	\item[(i)] $X$ is regularly varying with index $\alpha$.
  	\item[(ii)] There exists an $E$-valued time series
  	  $\{\bm Y_t\}_{t \in \ZZ}$ with 
  	  $\PP(\|\bm Y_0\|_\infty > y) = y^{-\alpha}$ for $y \geq 1$,
  	  called the \emph{tail process}, such that, for every finite $T \subset \ZZ$,
  	  $$ \calL(\{x^{-1} \bm X_t\}_{t \in T} \mid \|\bm X_0\|_\infty > x)
  	     \stackrel{x \to \infty}{\longrightarrow}
  	     \calL(\{\bm Y_t\}_{t \in T})  \quad \text{weakly in } E^T.$$
  	\item[(iii)] There exists an $E$-valued time series
  	  $\{\bm \Theta_t\}_{t \in \ZZ}$ with $\|\bm \Theta\|_\infty= 1$ a.s., called the \emph{spectral tail process},
  	  such that, for every finite $T \subset \ZZ$,
  	  	  $$ \calL(\{\|\bm X_0\|_\infty^{-1} \bm X_t\}_{t \in T} \mid \|\bm X_0\|_\infty > x)
  	     \stackrel{x \to \infty}{\longrightarrow}
  	     \calL(\{\bm \Theta_t\}_{t \in T})  \quad \text{weakly in } E^T.$$  
  \end{enumerate}   	
The tail process and the spectral tail process are related by 
$\calL(\{\bm Y_t\}_{t \in \ZZ}) =
  \calL(\{P \bm \Theta_t\}_{t \in \ZZ})$,	
where $P$ is an $\alpha$-Pareto variable independent of the process $\{\bm \Theta_t\}_{t \in \ZZ}$. 
\end{thm}

Note that, even if $\bm X_t$ is not univariate, but is a regularly varying multivariate or functional time series, we can transform it to a univariate time series by taking norms. Then, by definition, $\{\|\bm X_t\|_\infty\}_{t \in \ZZ}$ is a regularly varying univariate time series with tail process $\{\|\bm Y_t\|_\infty\}_{t \in \ZZ}$ and spectral tail process $\{\|\bm \Theta_t\|_\infty\}_{t \in \ZZ}$. Thus, applying the norm allows for the application of classical tools for the analysis of extremes in regularly varying time series. In practice, however, extremes are not only defined via the norm, but in terms of more general \emph{risk functionals}, see also \citet{dombry-ribatet-15}, \citet{thibaud-opitz-15} and \citet{defondeville-davison-18}. Thus, in the following, we analyze the extremal behavior of the time series with respect to such a functional.


\section{Clusters of extremes with respect to a homogeneous risk functional}
\label{sec:risk-fun-regvar}

\subsection{Risk functionals and regular variation}

Building upon the background theory laid out in Section~\ref{sec:regvar}, we study the functional time series $\{\bm X_t\}_{t \in \ZZ}$ and consider extremes defined in terms of threshold exceedances of a so-called risk functional, i.e., a functional 
$r: C_+(S) \to [0,\infty)$
that is continuous and positively $1$-homogeneous, i.e.,
$$ r(cf) = c r(f), \quad c \geq 0, \ f \in C_+(S).$$
As a first result, we generalize the fact that $\{\|\bm X_t\|_\infty\}_{t \in \ZZ}$ is a regularly varying univariate time series and state that the same holds true for any time
series of the type $\{r(\bm X_t)\}_{t \in \ZZ}$. This statement is closely related to results in \citet{dombry-ribatet-15} and \citet{dombry-hashorva-soulier-18}.
For completeness, we provide a short proof in the Supplementary Material.

\begin{prop} \label{prop:risk-fun}
  Let $\{\bm X_t\}_{t \in \ZZ}$ be a stationary $C_+(S)$-valued time series
  that is regularly varying with index $\alpha > 0$ and spectral tail process
  $\{\bm \Theta_t\}_{t \in \ZZ}$, and  let $r: C_+(S) \to [0,\infty)$ be a risk functional as defined above. Then, the univariate time series
  $\{r(\bm X_t)\}_{t \in \ZZ}$ is regularly varying with index $\alpha$ and
  spectral tail process $\{\bm \Theta_t^{r}\}_{t \in \ZZ}$ whose law is given by
  $$ \PP(\{\bm \Theta^{r}_t\}_{t \in \ZZ} \in A) = \frac 1 {\EE\{r(\bm \Theta_0)^\alpha\}}
     \int_{C_+(S)^\ZZ} \mathbf{1}\left\{  \left\{ \frac{r(\bm \theta_t)}{r(\bm \theta_0)}\right\}_{t \in \ZZ} \in A \right\} r(\bm \theta_0)^\alpha \, \PP_{\bm \Theta}(\mathrm{d} \bm \theta) $$
  for any measurable set $A \subset C_+(S)^\ZZ$. In particular, $r(\bm \Theta_0^r) = 1$ a.s..
\end{prop}

\subsection{Clusters of extremes} \label{sec:clustersize}

Denoting a $C_+(S)$-valued random object $\bm X_t$ at some fixed time $t \in \ZZ$ as an extreme observation if and only if $r(\bm X_t) > u$ for some large threshold $u$, we can then follow the notation of \citet{oesting-schnurr-19} and call an
$\ell$-dimensional vector of processes  $(\bm X_i)_{i=t}^{t+\ell-1}$ a $u$-exceedance cluster of size $\ell \in \NN$ with respect to $r$ if and only if 
$$ r(\bm X_{t-1}) \leq u, r(\bm X_t) > u, \ldots, r(\bm X_{t+\ell-1}) > u \ \text{and} \ r(\bm X_{t+\ell}) \leq u.$$
A natural interpretation of these clusters is that all the data points within a cluster
form an extreme event (such as a severe precipitation event spanning several time points, or a high sea temperature persisting over consecutive days), while different extreme events are separated by at least one non-exceedance of the prescribed threshold $u$. This definition naturally gives rise to questions that may be potentially interesting from a scientific perspective, such as ``How long does an extreme event last?'', or in other words, ``What is the size of a cluster of $u$-exceedances with respect to $r$?''. To answer such questions, we need to characterize the distribution of the size $C_u^{r}$ of a randomly selected $u$-exceedance cluster from the time series $\{\bm X_t\}_{t \in \ZZ}$ with respect to  $r$. The cluster size distribution may be defined as
\begin{align*}
 \PP(C_u^{r} = \ell) ={}& \frac{\PP(\text{there is a cluster of $u$-exceedances of size $\ell$ starting at $t=0$})}{\PP(\text{there is a cluster of $u$-exceedances starting at $t=0$} )} \\ 
 ={}& \frac{\PP(r(\bm X_{-1}) \leq u, r(\bm X_0) > u, \ldots, r(\bm X_{\ell-1}) > u, r(\bm X_\ell) \leq u)}{\PP(r(\bm X_{-1}) \leq u, r(\bm X_0) > u)}, \qquad \ell \in \NN.
\end{align*}

Analogously to \citet{oesting-schnurr-19}, regular variation of
$\{r(\bm X_t)\}_{t\in \ZZ}$ stated in Prop.~\ref{prop:risk-fun} implies the existence of a limit distribution as $u \to \infty$, namely,
\begin{align}
\lim_{u \to \infty} \PP(C_u^{r} = \ell) 
={}& \frac{\PP(P \cdot r(\bm \Theta^r_{-1}) \leq 1, \, P \cdot r(\bm \Theta^r_0) > 1, \,
  	 \ldots, \, P \cdot r(\bm \Theta^r_{\ell-1}) > 1, \, P \cdot r(\bm \Theta^r_\ell) \leq 1)}
     {\PP(P \cdot r(\bm \Theta^r_{-1}) \leq 1, \, P \cdot r(\bm \Theta^r_0) > 1)} 
     \nonumber \\
={}& \frac{ \PP( \max\{ 1/r(\bm \Theta^r_0), \ldots, 1/r(\bm \Theta^r_{\ell-1})\} 
            < P \leq \min\{1/r(\bm \Theta^r_{-1}), 1/r(\bm \Theta^r_\ell)\}) }
          {\PP( 1/r(\bm \Theta^r_0) < P \leq 1/r(\bm \Theta^r_{-1})) } \nonumber \\
={}& \frac{\EE\left[ \min\{r(\bm \Theta_0^r)^\alpha, \ldots, r(\bm \Theta_{\ell-1}^r)^\alpha\}  - \max\{r(\bm \Theta_{-1}^r)^\alpha, r(\bm \Theta_\ell^r)^\alpha\} \right]_+}{\EE\left[r(\bm \Theta_{0}^r) - r(\bm \Theta_{-1}^r)^\alpha \right]_+} \nonumber \\
={}& \frac{\EE\left[ \min\{1, r(\bm \Theta_1^r)^\alpha, \ldots  \wedge r(\bm \Theta_{\ell-1}^r)^\alpha\}  - \max\{r(\bm \Theta_{-1}^r)^\alpha, r(\bm \Theta_\ell^r)^\alpha\} \right]_+}{\EE\left[1 - r(\bm \Theta_{-1}^r)^\alpha \right]_+},
\label{eq:lim-cluster-size}
\end{align}     
for $l \in \NN$, where $P$ is an $\alpha$-Pareto random variable independent of the spectral tail process
$\{\bm \Theta_t^r\}_{t\in\ZZ}$. Note that, for the last equality, we used that $r(\bm \Theta_{0}^r)=1$ a.s..

\subsection{Extension to the non-heavy-tailed case} \label{subsec:regvar-extension}

Remind that, for the definition of clusters, we required the $C_+(S)$-valued time series $\{\bm X_t\}_{t \in \ZZ}$ to be regularly varying. \citet{segers-zhao-meinguet-2017} showed that this holds if and only if it is regularly varying as a random element of the metric space $C_+(S)^{\ZZ}$. Thus, by results such as Thm-~2.4 in \citet{lin-dehaan-2001} or Lemma 2.2 in \cite{davis-mikosch-08}, $\{\bm X_t\}_{t \in \ZZ}$ is regularly varying with index $\alpha$ if and only if $\{\bm X_t\}_{t \in \ZZ}$ is in the max-domain of attraction of a max-stable process $Z=\{\bm Z_t\}_{t \in \ZZ}$ with $\alpha$-Fr\'echet margins, i.e., 
$a_n^{-1} \max_{i=1,\ldots,n} X^{(i)} \to_d Z$ 
as processes in $C_+(S)^\ZZ$, where $X^{(1)}, X^{(2)}, \ldots$ are independent copies of $X$ and $a_n$ are constants as in Def.~\ref{def:regvar}. 
Thus, we can extend our results to stationary $C(S)$-valued time series $\{\bm W_t\}_{t \in \ZZ}$ that can be monotonically transformed to a regularly varying $C_+(S)$-valued time series $\{\bm X_t\}_{t \in \ZZ}$, i.e.,
\begin{equation} \label{eq:xy-trafo}
  X(\bm s,t) = \phi(W(\bm s,t)), \quad \bm s \in S, \ t \in \ZZ,
\end{equation}
where $\phi: (-\infty, w^*) \to [0,\infty)$ is some nondecreasing bijective marginal transformation and $w^* \in \RR \cup \{\infty\}$ is the (joint) upper endpoint of the marginal distributions of $W^*$.

Now, assume that the risk functional $r$ satisfies
\begin{align} \label{eq:commute-r}
 \phi(r(f)) = r(\phi(f)) 
\end{align}
for any continuous function $f: S \to (-\infty,y^*)$. For instance, irrespective of the choice of $\phi$, Condition \eqref{eq:commute-r} holds true for the following specific examples:
\begin{itemize}
	\item the spatial minimum/maximum : $r(f) = \min_{\bm s \in S} f(\bm s)$ and $r(f) = \max_{\bm s \in S} f(\bm s)$,
	\item spatial quantiles:
	  $$ r(f) = \inf\left\{ q \in \RR: \, |S|^{-1}\int_S \one\{f(\bm s) \geq q\} \, \mathrm{d}\bm s > p  \right\}, \qquad p \in (0,1), \quad |S|=\int_S 1 \, \mathrm{d}\bm s. $$
\end{itemize}

Under Assumptions \eqref{eq:xy-trafo} and \eqref{eq:commute-r} it is reasonable to consider $\phi^{-1}(u)$-exceedance clusters in $\{\bm W_t\}_{t \in \ZZ}$ with respect to $r$ for $u \to \infty$, that is, as $\phi^{-1}(u) \to w^*$. The probability that a randomly selected cluster of exceedances is of size $\ell$ then equals 
\begin{align*}
& \frac{\PP(r(\bm W_{-1}) \leq \phi^{-1}(u), r(\bm W_0) > \phi^{-1}(u), \ldots, r(\bm W_{\ell-1}) > \phi^{-1}(u), r(\bm W_\ell) \leq \phi^{-1}(u))}{\PP(r(\bm W_{-1}) \leq \phi^{-1}(u), r(\bm W_0) > \phi^{-1}(u))} \\
={}& \frac{\PP(r(\bm X_{-1}) \leq u, r(\bm X_0) > u, \ldots, 
	r(\bm X_{\ell-1}) > u, r(\bm X_\ell) \leq u)}{\PP(r(\bm X_{-1}) \leq u, r(\bm X_0) > u)}.	
\end{align*}
and, by \eqref{eq:lim-cluster-size}, the limit for the right-hand side exists as $X=\phi(W)$ is regularly varying.

Examples of such processes $\{\bm W_t\}_{t \in \ZZ} = \{W(\bm s,t)\}_{\bm s \in S, t \in \ZZ}$ 
comprise all processes with tail-equivalent marginal distributions that are in the max-domain of attraction of an arbitrary max-stable process $Z$, denoted by $W \in {\rm MDA}(Z)$. This is due to the fact that $Y^* \in {\rm MDA}(Z)$ if and only if the following two conditions hold \citep[see][Corollary 2.2]{aulbach-falk-hofmann-zott-2015}:
\begin{itemize}
	\item $W(\bm s,t) \in {\rm MDA}(Z(\bm s,t))$ for all fixed $\bm s \in S$, $t \in \ZZ$,
	\item and the copula process of $W$ is in the max-domain of attraction of a marginally transformed version of $Z$ with reversed Weibull margins.
\end{itemize}
In other words, $\phi(W)$ is thus still in the max-domain of attraction of some max-stable process as long as the marginal distributions are attracted by a generalized extreme value distribution. In particular, if the marginal distributions of $X=\phi(W)$ are in the max-domain of attraction of an $\alpha$-Fr\'echet distribution, $\{\bm X_t\}_{t \in \ZZ}$ is regularly varying as discussed above.

\section{Within-cluster behavior of extremes} \label{sec:inside-cluster}

\subsection{General setting} \label{sec:gen-setting}
In the following sections, we consider a regularly varying time series
$\{\bm X_t\}_{t \in \ZZ}$, and take a closer look at the behavior of the time series
within a cluster of exceedances. More precisely, we shall study the behavior of the vector of normalized processes $(u^{-1}\bm X_i)_{i=t}^{t+\ell-1}$ provided that $(\bm X_i)_{i=t}^{t+\ell-1}$ is a cluster of $u$-exceedances with respect to a risk functional $r$, i.e., we condition on the event $r(\bm X_{t-1}) \leq u$, $r(\bm X_t) > u$, $\ldots$, $r(\bm X_{t+\ell-1}) > u$ and $r(\bm X_{t+\ell}) \leq u$.

While regular variation guarantees the existence of a limit expression for
$$ \PP( (u^{-1} \bm X_t, \ldots, u^{-1} \bm X_{t+\ell-1}) \in A \mid r(\bm X_{t-1}) \leq u, r(\bm X_t) > u, \ldots, r(\bm X_{t+\ell-1}) > u, r(\bm X_{t+\ell}) \leq u) $$
as $u \to \infty$ whenever 
$\PP( (P\bm\Theta_0,\ldots,P\bm\Theta_{\ell-1}) \in \partial A) = 0$ for a broad range of Borel sets $A \subset (C_+(S))^\ell$, we shall henceforth study the limiting within-cluster behavior of certain functionals applied separately to each of the spatial fields $\bm X_t, \ldots, \bm X_{t+\ell-1}$. More precisely, we here consider three types of functionals, which have a practical interest:
\begin{itemize}
	\item the risk functional $r$ itself (Section~\ref{sec:lim-riskfun}), which provides insights into the (evolving) severity of an extreme event during a cluster of exceedances.
	\item spatial risk measures (Section~\ref{sec:lim-spatrisk}), which provide insights into the (evolving) spatial extent of an extreme event during a cluster of exceedances. 
	\item multivariate location measures (Section~\ref{sec:lim-loc}), which provide insights into the (evolving) position of an extreme event during a cluster of exceedances.
\end{itemize}

\subsection{Limit expressions for risk functionals} \label{sec:lim-riskfun}

As the risk functional $r$ is used to define extremes, it is natural to further analyze 
its behavior inside an extreme event, i.e., inside a cluster of $u$-exceedances with respect to
$r$. As the time series $\{r(\bm X_t)\}_{t \in \ZZ}$ is regularly varying by Prop.~\ref{prop:risk-fun}, we can exploit the results from \citet{oesting-schnurr-19} to derive the limiting distribution of the corresponding extremal ordinal patterns and other within-cluster characteristics.

Note that there are various definitions of $\ell$-ordinal patterns being related either on the rank \cite[e.g., in][]{bandt-2020} or the order \citep[e.g., in][]{keller-etal-07} of data. We here make use of the rank-based definition because we consider it more straightforward to interpret, thus deviating from the notion used in \citet{oesting-schnurr-19}, for instance, who propose to use ordinal patterns to describe stylized facts of extreme events.
More precisely, we define the $\ell$-ordinal pattern as the mapping $\Pi$ that maps
a vector $(X_i)_{i=1}^{\ell} \in \RR^\ell$ (without ties) to the unique permutation $\pi$ of $\{1,\ldots,\ell\}$ such that 
$$ X_{i} < X_{j}  \iff \pi(i) < \pi(j), $$
i.e., $\Pi(X)$ is the vector consisting of the ranks of the components of $X$.
For instance, whenever $X$ is itself some permutation of $\{1,\ldots,\ell\}$, the ordinal pattern $\Pi(X)$ equals $X$. Note that, in most of the applications we have in mind, ties occur very rarely. Thus, we do not consider this case further, but shall report any pattern including ties separately.

Denoting the probability that the $\ell$-ordinal pattern of $(r(\bm X_i))_{i=0}^{\ell-1}$
for a cluster of $u$-exceedances $(\bm X_i)_{i=0}^{\ell-1}$ equals a specific permutation
$\pi$ by $\PP_{u,\ell}(\pi)$, i.e.,
$$ \PP_{u,\ell}(\pi) = \PP( \Pi((r(\bm X_i))_{i=0}^{\ell-1}) = \pi \mid r(\bm X_{-1}) \leq u,
  r(\bm X_0) > u, \ldots, r(\bm X_{\ell-1}) > u, r(\bm X_\ell) \leq u), $$
we obtain the limit distribution
\begin{align}
 & \lim_{u \to \infty} \PP_{u,\ell}(\pi) \nonumber \\ 
 ={}& \frac{ \PP( \Pi((r(\bm \Theta_i^r))_{i=0}^{\ell-1}) = \pi, P \cdot r(\bm \Theta_{-1}^r) \leq 1, P \cdot r(\bm \Theta_{0}^r) > 1, \ldots, P \cdot r(\bm \Theta_{\ell-1}^r) > 1,
   P \cdot r(\bm \Theta_{l}^r) \leq 1)}
   { \PP(P \cdot r(\bm \Theta_{-1}^r) \leq 1, P \cdot r(\bm \Theta_{0}^r) > 1, \ldots,
   	   P \cdot r(\bm \Theta_{\ell-1}^r) > 1, P \cdot r(\bm \Theta_{\ell}^r) \leq 1)} \nonumber \\
  ={}& \frac{\EE[ (r(\bm \Theta_{0}^r)^\alpha \wedge \ldots \wedge r(\bm \Theta_{\ell-1}^r)^\alpha - r(\bm \Theta_{-1}^r)^\alpha \vee (r(\bm \Theta_{\ell}^r)^\alpha)_+  \bm 1\{\Pi((r(\bm \Theta_i^r))_{i=0}^{\ell-1}) = \pi\}]}
  {\EE(r(\bm \Theta_{0}^r)^\alpha \wedge \ldots \wedge r(\bm \Theta_{\ell-1}^r)^\alpha - r(\bm \Theta_{-1}^r)^\alpha \vee (r(\bm \Theta_{\ell}^r)^\alpha)_+}. 
  \label{eq:lim-ord-pattern}      
\end{align}      

\begin{rem}
	As the ordinal pattern is invariant under marginal transformations, by the same arguments as given in Section~\ref{subsec:regvar-extension}, the results above also 
	hold true more generally for marginally-transformed processes $W$ defined such that $\{\bm X_t\}_{t\in\ZZ} = \{\phi(\bm W_t)\}_{t\in\ZZ}$ is regularly varying as in \eqref{eq:xy-trafo}, provided that \eqref{eq:commute-r} holds.
\end{rem}

Even more generally, it follows directly from Prop.~\ref{prop:risk-fun} that
\begin{align*}
 & \lim_{u \to \infty} \PP( (u^{-1} r(\bm X_i))_{i=0}^{\ell-1} \in A \mid 
     r(\bm X_{-1}) \leq u, r(\bm X_0) > u, \ldots, r(\bm X_{\ell-1}) > u, r(\bm X_\ell) \leq u)\\
={}& \frac{\PP( (P \cdot r(\bm \Theta^r_i))_{i=0}^{\ell-1} \in A, 
	   P \cdot r(\bm \Theta_{-1}^r) \leq 1, P \cdot r(\bm \Theta_{0}^r) > 1, \ldots,
	P \cdot r(\bm \Theta_{\ell-1}^r) > 1, P \cdot r(\bm \Theta_{\ell}^r) \leq 1)} 
    {\PP(P \cdot r(\bm \Theta_{-1}^r) \leq 1, P \cdot r(\bm \Theta_{0}^r) > 1, \ldots,
    	P \cdot r(\bm \Theta_{\ell-1}^r) > 1, P \cdot r(\bm \Theta_{\ell}^r) \leq 1)}
\end{align*}
for all measurable $A \in (0,\infty)^\ell$ satisfying 
$\PP( (P \cdot r(\bm \Theta^r_i))_{i=0}^{\ell-1} \in \partial A) = 0$. However, in this work, we mainly focus on ordinal patterns.

\subsection{Limit expressions for spatial risk measures}  \label{sec:lim-spatrisk}

Besides the behavior of the risk functional itself inside a cluster, there are also other quantities of practial interest, such as the spatial extent of an extreme event for instance. To this end, \citet{koch-17} proposed to consider the normalized
spatially aggregated ``loss'' due to a (random) function $f \in C_+(S)$ defined by
$$ L_u(f) = \frac {1}{|S|} \int_S E(\bm s) \one\{f(\bm s) > u\}\,\mathrm{d}\bm s,$$
where $E$ is a positive \emph{exposure function}. Here, it is important to note that, due to the indicator function, $L_u$ is not homogeneous and thus not a valid risk functional.

In the special case where $E \equiv 1$, $L_u$ corresponds to the percentage of
the area $S$ that is affected by an extreme event defined as a pointwise exceedance of the threshold $u$. We are interested in the joint limiting behavior of the spatially aggregated ``losses'' of $\bm X_t,\ldots,\bm X_{t+\ell-1}$ provided that $(\bm X_i)_{i=t}^{t+\ell-1}$ forms a cluster of extremes. More generally, we can consider the functional $m: \mathrm{Bin}(S) \to [0,\infty]$
where $\mathrm{Bin}(S) = \{f: S \to \{0,1\}, \ f \text{ measurable}\}$ is the space of
measurable binary functions on $S$. For instance, choosing  
$$ m(f) = |S|^{-1} \int_S E(\bm s) \cdot f(\bm s) \, \mathrm{d}\bm s,$$
 we can rewrite $L_u(\bm X_t) = m(\one\{u^{-1} \bm X_t > 1\})$.

In the following, with a slight abuse of notation, we write $m(A)$ for $m(\one\{A\})$. Furthermore, we assume that $m(\bm 0_S)=0$ where $\bm 0_S$ denotes the zero function on $S$. Under these assumption, the following result which is proven in the Supplementary Material, on the limit distribution of spatial risk measures can be stated.

\begin{prop} \label{prop:spat-risk}
 Let $\{\bm X_t\}_{t \in \ZZ}$ be a stationary $C_+(S)$-valued time series that
 is regularly varying with index $\alpha>0$ and spectral tail process $\{\bm \Theta_t\}_{t \in \ZZ}$, and let $m: \mathrm{Bin}(S) \to [0,\infty)$ be such that $m(\bm 0_S) =0$. 
 If the joint distribution of $(m(\{\bm \Theta_0>\eta\}), \ldots, 
 {m(\{\bm \Theta_{\ell-1}>\eta\})})$ depends continuously on $\eta \in [0,1]$ for each $\ell >0$, 
 then, for any measurable set $A \subset (0,\infty)^{\ell}$ with
 $\PP\left( (m(\{\bm \Theta_0>\eta\}), \ldots,
 m(\{\bm \Theta_{\ell-1}>\eta\})) \in \partial A\right) = 0$
 for a.e.\ $\eta \in [0,1]$, we have that
 \begin{align*}
 & \PP\big( 
   (m(\{\bm X_0>u\}),\ldots,m(\{\bm X_{\ell-1}>u\})) \in A 
 \mid \\
 & \qquad \qquad r(\bm X_{-1}) \leq u, r(\bm X_0) > u, \ldots, r(\bm X_{\ell-1}) > u, r(\bm X_\ell) \leq u\big) \\
 \stackrel{u \to \infty}{\longrightarrow}{}&
 \left(\EE[\min\{1, r(\bm \Theta_{1})^\alpha, \ldots, r(\bm \Theta_{\ell-1})^\alpha\} 
     - \max\{r(\bm \Theta_{-1})^\alpha,r(\bm \Theta_\ell)^\alpha\}]_+\right)^{-1}\\
  & \quad    \cdot 
      \int_0^1 \PP\big( (m(\{\bm \Theta_0^{\alpha}>\eta\}), \ldots,
    m(\{\bm \Theta_{\ell-1}^{\alpha}>\eta\})) \in A,\\
    & \qquad \qquad \qquad r(\bm \Theta_{-1})^{\alpha} \leq \eta, \, r(\bm \Theta_0)^{\alpha} > \eta, \ldots, r(\bm \Theta_{\ell-1})^{\alpha} > \eta, \,
    r(\bm \Theta_\ell)^{\alpha} \leq \eta \big) \, \mathrm{d}\eta.
 \end{align*}
\end{prop}

As an example, similarly to Section~\ref{prop:risk-fun}, we can consider ordinal patterns, but now applied to the spatial risk measure $m$ rather than the risk functional $r$. Specifically, by a simple application of the results above, we can find the asymptotic distribution of 
$$ \Pi((m(\{\bm X_i > u\}))_{i=0}^{\ell-1}) \mid r(\bm X_{-1}) \leq u, r(\bm X_0) > u, \ldots, r(\bm X_{l-1}) > u, r(\bm X_l) \leq u),$$
as $u \to \infty$. These ordinal patterns provide important information about the evolution of the spatial extent (i.e., spatial risk) of an extreme event of a certain duration. 

\begin{rem}
	For a process $W$ related to $X$ via $\bm X_t = \phi(\bm W_t)$ for all $t \in \ZZ$ as in \eqref{eq:xy-trafo},
	we can directly rewrite $m(\{\bm X_t > u\}) = m(\{\bm W_t > \phi^{-1}(u)\})$.
	Thus, if \eqref{eq:commute-r} holds, we have that 
	\begin{align*}
	& \PP\big( 
	(m(\{\bm Y_0 > \phi^{-1}(u)\}),\ldots,m(\{\bm Y_{l-1}>\phi^{-1}(u)\})) \in A 
	\mid \\
	& \hspace{3cm} r(\bm Y_{-1}) \leq \phi^{-1}(u), r(\bm Y_0) > \phi^{-1}(u), \ldots, r(\bm Y_{\ell-1}) > \phi^{-1}(u), r(\bm Y_l) \leq \phi^{-1}(u)\big) \\
	={} & \PP\big( 
	(m(\{\bm X_0>u\}),\ldots,m(\{\bm X_{\ell-1}>u\})) \in A 
	\mid \\
 & \hspace{3cm}  r(\bm X_{-1}) \leq u, r(\bm X_0) > u, \ldots, r(\bm X_{\ell-1}) > u, r(\bm X_\ell) \leq u\big) 
	\end{align*}
	and Prop.~\ref{prop:spat-risk} applies if $\{\bm X_t\}_{t \in \ZZ}$ is regularly varying.
\end{rem}

\subsection{Limit expressions for location measures} \label{sec:lim-loc}
Beyond the intensity and the spatial extent, we may be interested to infer whether an extreme event moves over space, i.e., whether (and how) its position changes during a cluster of exceedances. To tackle this question, we consider general functionals $c: C_+(S) \to S$ describing the location of an extreme event. Having in mind the interpretation as the ``center'' of the area affected by the extreme event, we want to allow for cases where this location depends on the area where the threshold (here, set to $1$ for simplicity) is exceeded. Furthermore, once this area is fixed, we want the location to be invariant under scalar multiplication, i.e., we assume the existence of some functional $\tilde c: C_+(S) \times \mathrm{Bin}(S) \to S$ such that
$$ c(f) = \tilde c \left( \frac{f}{r(f)}, \one\{f > 1\} \right)
  =: \tilde c \left( \frac{f}{r(f)}, \{f > 1\} \right),$$
and certain continuity conditions specified below are satisfied. 
Typical examples of such functionals include:
\begin{itemize}
 \item the peak location,
      $c(f) = \argmax_{\bm s \in S} f(\bm s)$; 
 \item the centroid of the region exceeding the threshold,
 $$ c(f) = \frac{1}{\int_{\{\bm s\in S: f(\bm s) > 1 \}} 1 \, \mathrm{d} \bm s} \int_{\{\bm s\in S: f(\bm s) > 1\}} \bm s \, \mathrm{d} \bm s;$$
 where the second integral is to be interpreted componentwise;
 \item the centroid weighted by the intensity, $f(\bm s)$, of the event itself,
    $$ c(f) =\frac{1}{\int_{S} f(\bm s) \, \mathrm{d} \bm s} \int_{S} \bm s f(\bm s) \, \mathrm{d} \bm s = \begin{pmatrix} 
        \frac{\int_{S} s_1 f(\bm s) \, \mathrm{d}\bm s}{\int_{S} f(\bm s) \, \mathrm{d}\bm s}  \\
        \vdots \\
        \frac{\int_{S} s_d f(\bm s) \, \mathrm{d}\bm s}{\int_{S} f(\bm s) \, \mathrm{d}\bm s}  
	    \end{pmatrix};$$
 \item the location defined by the componentwise median of the coordinates of all points exceeding the threshold, 
 $$c(f)={\rm median}\{\bm s\in S:f(\bm s)>1\}.$$
\end{itemize}

We are interested in the behavior of $c(u^{-1} \bm X_t)$ if $\bm X_t$ is extreme in the sense that $r(\bm X_t)>u$ for some large threshold $u$ over consecutive time points. The following proposition, proven in the Supplementary Material, characterizes the limiting distribution of such an event.

\begin{prop} \label{prop:location}
  Let $\{\bm X_t\}_{t \in \ZZ}$ be a stationary $C_+(S)$-valued time series  
  that is regularly varying with index $\alpha > 0$ and spectral tail process
  $\{\bm \Theta_t\}_{t \in \ZZ}$. Let $r: C_+(S) \to [0,\infty)$
  be a continuous positively $1$-homogeneous risk functional and let $c: C_+(S) \to S$ be of the form 
  $$ c(f) = \tilde c \left( \frac{f}{r(f)}, \one\{f > 1\} \right),$$
  for some functional $\tilde c: C_+(S) \times \mathrm{Bin}(S) \to S$
  where we formally set $\tilde c(\cdot, \bm 0_S) = \bm \infty.$
  Moreover, assume that, for $\ell>0$, the joint distribution of $(c(\eta^{-1} \bm \Theta_0), \ldots, c(\eta^{-1} \bm \Theta_{\ell-1}))$ depends continuously on $\eta \in (0,1]$.
  Then, for any $\ell \in \NN$ and any measurable set $A \subset S^{\ell}$ satisfying
  $$ \PP( (c(\bm \Theta_0), \ldots, c(\bm \Theta_{\ell-1})) \in \partial A) = 0,$$
  we have that
  \begin{align*}
   \PP( (c&(u^{-1}\bm X_0), \ldots, c(u^{-1}\bm X_{\ell-1})) \in A \mid 
    r(\bm X_{-1}) \leq u, r(\bm X_0) > u, \ldots, r(\bm X_{\ell-1}) > u, r(\bm X_\ell) \leq u)\\ 
   \stackrel{u \to \infty}{\longrightarrow}{}&
  \left[\EE(\min\{1, r(\bm\Theta_1)^\alpha, \ldots, r(\bm\Theta_{l-1})^\alpha\}
  	- \max\{r(\bm\Theta_{-1})^\alpha, r(\bm\Theta_\ell)^\alpha\})_+\right]^{-1} \\
   & \qquad \cdot \int_0^1 \PP\big( (\tilde c(\bm \Theta_0, \{\bm \Theta_0^\alpha > \eta\}), \ldots, \tilde c(\bm \Theta_{\ell-1}, \{\bm \Theta_{\ell-1}^\alpha > \eta\})) \in A,\\
& \qquad \qquad \qquad \quad r(\bm \Theta_{-1})^\alpha \leq \eta, r(\bm \Theta_0)^\alpha > \eta, \ldots, r(\bm \Theta_{\ell-1})^\alpha > \eta, r(\bm \Theta_\ell)^\alpha \leq \eta \big)
\, \mathrm{d}\eta.
  \end{align*}
\end{prop}

\begin{rem}
	Note that, by the same arguments as given in Section~\ref{subsec:regvar-extension}, the results above also hold true for marginally-transformed processes $W$ such that $\{\bm X_t\}_{t\in\ZZ} = \{\phi(\bm W_t)\}_{t\in\ZZ}$ is regularly varying as in \eqref{eq:xy-trafo}, provided that $c$ is invariant under the transformation $\phi$, i.e.,
	$ c(\phi(f)) = c(f) $
	for any continuous function $f: S \to (-\infty,y^*)$. This is true for the peak location functional, the centroid of the region exceeding the threshold, and for the componentwise median location of threshold exceedances, but not for the weighted centroid.
\end{rem}

\section{Inference for limit distributions} \label{sec:estim}

Even though Equation \eqref{eq:lim-cluster-size}, Equation \eqref{eq:lim-ord-pattern},
Prop.~\ref{prop:spat-risk} and Prop.~\ref{prop:location} provide explicit
expressions for some specific cases in terms of the spectral tail process $\{\bm \Theta_t\}_{t \in \ZZ}$, all the limit expressions considered in Section
\ref{sec:risk-fun-regvar} and Section~\ref{sec:inside-cluster} are of the form
\begin{align*}
 \lim_{u \to \infty} \frac{\PP((u^{-1} \bm X_i)_{i=-1}^\ell \in A)}{\PP((u^{-1} \bm X_i)_{i=-1}^\ell \in A_0)} ={}& \frac{\PP((P \cdot \bm \Theta_i)_{i=-1}^\ell \in A)}{\PP((P \cdot \bm \Theta_i)_{i=-1}^\ell \in A_0)}
 ={} \frac{\PP((\bm Y_i)_{i=-1}^\ell \in A)}{\PP((\bm Y_i)_{i=-1}^\ell \in A_0)}
 {}={} \frac{\mu_{\{-1,\ldots,\ell\}}(A)}{\mu_{\{-1,\ldots,\ell\}}(A_0)},
\end{align*}
where $\{\bm Y_t\}_{t \in \ZZ}$ and $\mu_{T}$, $T \subset \ZZ$, denote the tail process
and the exponent measure, given in Def.~\ref{def:regvar} and Thm.~\ref{thm:char-regvar}, respectively. In all of the cases considered so far, the sets $A_0$ and $A$ include the condition that an extreme event occurs at time $t=0$, i.e.,
$$ A_0, A \subset \{(\bm X_{-1}, \ldots, \bm X_\ell) \in (C_+(S))^{\ell+2}: \ r(\bm X_0) > 1\}. $$
Given observations $\bm X_{-1}, \ldots, \bm X_n \in C_+(S)$, a natural estimator for the desired limiting expression $\mu_{\{-1,\ldots,\ell\}}(A) / \mu_{\{-1,\ldots,\ell\}}(A_0)$ is the
ratio estimator
\begin{equation} \label{eq:ratio-estimator}
\widehat{R}_{n,u}(A,A_0) 
= \frac{\sum_{k=0}^{n-\ell} \mathbf{1}\{ (u^{-1} \bm X_i)_{i=k-1}^{k+\ell} \in A\}}
       {\sum_{k=0}^{n-\ell} \mathbf{1}\{ (u^{-1} \bm X_i)_{i=k-1}^{k+\ell} \in A_0\}}
       =: \frac{\widehat{\PP}_{n,u}(A)}{\widehat{\PP}_{n,u}(A_0)}.
\end{equation}      
We shall impose certain conditions on the time series $\{\bm X_t\}_{t \in \ZZ}$ to ensure asymptotic normality of the estimator. More precisely, we follow \cite{oesting-schnurr-19}
who adapt conditions used in \cite{davis-mikosch-09} for the estimation of the extremogram, in order to show asymptotic normality for ratio estimators of the same type as in \eqref{eq:ratio-estimator} in the context of a regularly varying univariate time series. We here extend these results to the functional time series setting. 
We impose weak (and eventually vanishing) long-range temporal dependence. Specifically, consider the $\alpha$-mixing coefficient
\begin{align} \label{eq:def-alpha}
	\alpha_h ={}&  \sup_{A,B \in \calB(C_+(S)^\NN)} |\PP(\{ \bm X_t\}_{t \leq 0} \in A, \, \{\bm X_t\}_{t \geq h} \in B) - \PP(\{\bm X_t\}_{t \leq 0} \in A) \PP(\{\bm X_t\}_{t \geq h} \in B)|,
\end{align}
where $h \in \NN_0$. We assume the following mixing and anti-clustering condition:
\begin{condition}[M]
There exist a sequence $\{u_n\}_{n \in \NN} \subset \RR$ of thresholds and 
an intermediate sequence
$\{r_n\}_{n \in \NN} \subset \NN$ 
with $\lim_{n \to \infty} u_n = \lim_{n \to \infty} r_n = \infty$,
$ \lim_{n \to \infty} n \PP(\|\bm X_0\|_\infty > u_n) = \infty$, 
$ \lim_{n \to \infty} r_n \PP(\|\bm X_0\|_\infty > u_n) = 0$ such that
\begin{equation} \label{eq:mixing}
	\lim_{n \to \infty} \frac 1 {\PP(\|\bm X_0\|_\infty > u_n)} 
	\sum\nolimits_{h=r_n}^\infty \alpha_h = 0
\end{equation}
and, for all $\varepsilon > 0$,
\begin{equation} \label{eq:anticlustering}
	\lim_{k \to \infty} \limsup_{n \to \infty} \sum\nolimits_{h=k}^{r_n} \PP(\|\bm X_h\|_\infty > \varepsilon u_n \mid \|\bm X_0\|_\infty > \varepsilon u_n) = 0.
\end{equation}
\end{condition}

We can then show the following result that
guarantees asymptotic normality of our estimator, based on a proof analoguous to the results of \cite{oesting-schnurr-19}:

\begin{prop} \label{prop:asympt-normal}
Let $\{\bm X_t\}_{t \in \ZZ}$ be a stationary $C_+(S)$-valued time series that is regularly varying with index $\alpha > 0$ and tail process $\{\bm Y_t\}_{t \in \ZZ}$ whose finite-dimensional distributions are 
denoted by $(\mu_I)_{I \subset \ZZ}$. Moreover, let $A, A_0  \subset C_+(S)^{\ell+1}$ be  continuity sets with respect to 
$\mu_{\{-1,\ldots,\ell\}}$ bounded away from zero. We further assume that Condition (M) holds
and that the $\alpha$-mixing coefficients satisfy $\alpha_h \in \mathcal{O}(h^{-\delta})$ for some $\delta>2$. If the sequence of thresholds $\{u_n\}_{n \in \NN}$ additionally satisfies
\begin{equation} \label{eq:add-cond}
	\lim_{n \to \infty} n^{\delta/(4+\delta)} \PP(\|\bm X_0\|_\infty > u_n) = \infty,
\end{equation}
and
\begin{equation*}
	\lim_{n \to \infty} \sqrt{n\PP(\|\bm X_0\|_\infty > u_n)} \left[ \frac{\PP(\{\bm X_i\}_{i=-1}^\ell \in u_n A)}{\PP(\{\bm X_i\}_{i=-1}^\ell \in u_n A_0)} -  \frac{\mu_{\{-1,\ldots,\ell\}}(A)}{\mu_{\{-1,\ldots,\ell\}}(A_0)} \right] = 0,
\end{equation*}
then
$$ \sqrt{n \PP(\|\bm X_0\|_\infty > u_n)} \left( \widehat{R}_{n,u_n}(A,A_0) - \frac{\mu_{\{-1,\ldots,\ell\}}(A)}{\mu_{\{-1,\ldots,\ell\}}(A_0)} \right) $$
converges in distribution to a centered Gaussian random variable as $n \to \infty$.
\end{prop}

The asymptotic variance in Prop.~\ref{prop:asympt-normal} can be estimated via bootstrap techniques. Here, we make use of a multiplier block bootstrap which has proven to be appropriate for estimators of a similar type \citep{drees-15}. More precisely, we divide the time series data 
$\bm X_{-1}, \ldots, \bm X_n$ of length $n+1$ into $b_n$ data sets of length $\ell_n+2$, i.e.,
we consider blocks 
$$ (\bm X_t)_{t=-1}^{\ell_n},  (\bm X_t)_{t=\ell_n+1}^{2\ell_n+3}, \ldots,  (\bm X_t)_{t=n-\ell_n-1}^{n} $$
and denote the counterparts of $\widehat{\PP}_{n,u}(A)$ and $\widehat{\PP}_{n,u}(A_0)$ 
based on each of these $b_n$ blocks by
$$\widehat{\PP}^{(1)}_{\ell_n,u_n}(A), \ldots, \widehat{\PP}^{(b_n)}_{\ell_n,u_n}(A)\qquad \text{and}\qquad \widehat{\PP}^{(1)}_{\ell_n,u_n}(A_0), \ldots, \widehat{\PP}^{(b_n)}_{\ell_n,u_n}(A_0),$$
respectively. Then, a bootstrap sample of $\widehat{R}_{n,u_n}(A,A_0)$ can be obtained by repeatedly sampling i.i.d.\ random variables $\xi_1, \ldots, \xi_{b_n}$ with $\EE(\xi_j) = 0$ and $\Var(\xi_j)=1$ and considering
$$ \frac{ \sum_{j=1}^{b_n} (1 + \xi_j) \widehat{\PP}^{(j)}_{\ell_n,u_n}(A) }
        { \sum_{j=1}^{b_n} (1 + \xi_j) \widehat{\PP}^{(j)}_{\ell_n,u_n}(A_0) }. $$
We analyze the performance of this multiplier block bootstrap scheme numerically in our simulation study in Section~\ref{sec:simu}.

\section{Simulation study} \label{sec:simu}

\begin{figure}[t!]
    \centering
    \includegraphics[width=0.4\textwidth]{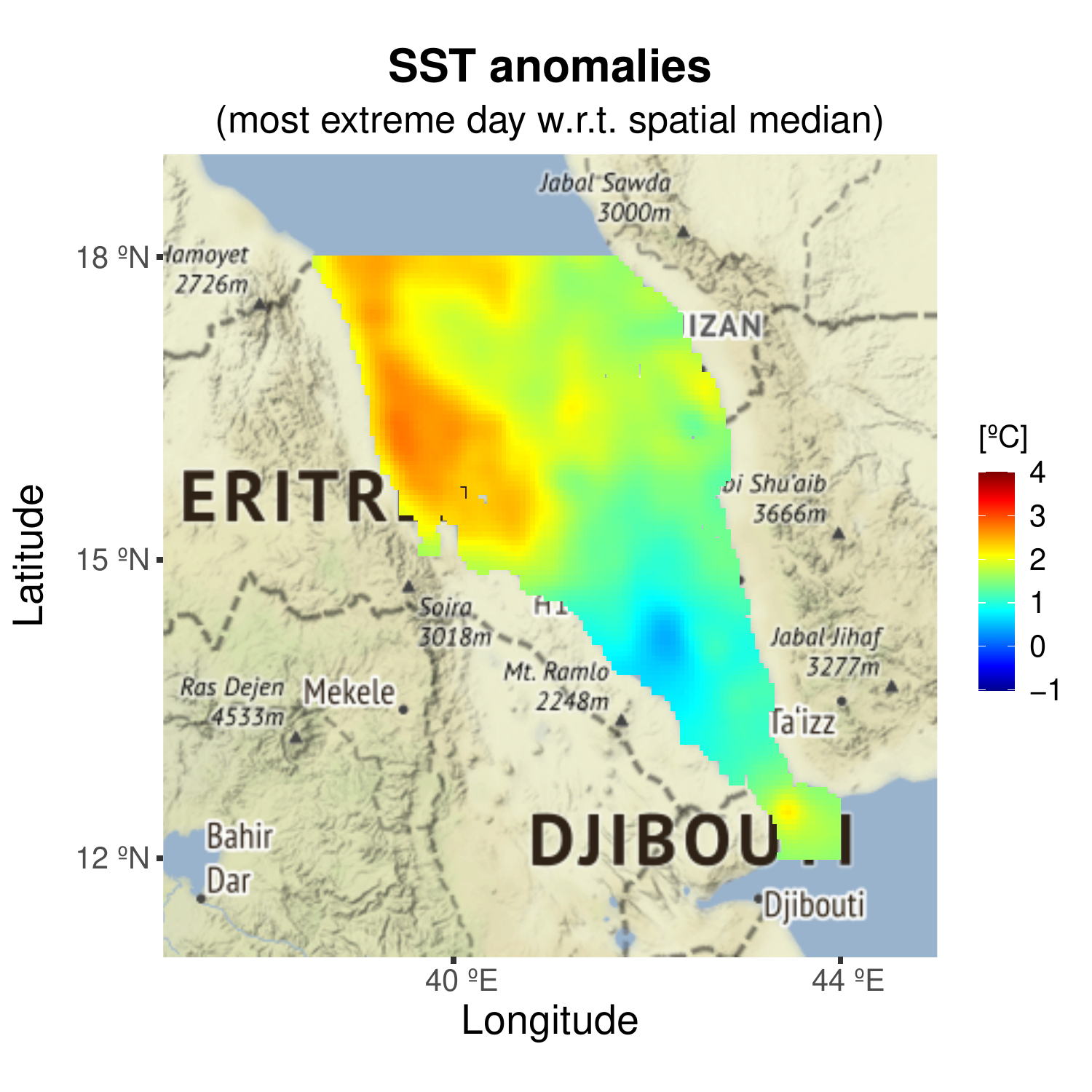} 
    \includegraphics[width=0.4\textwidth]{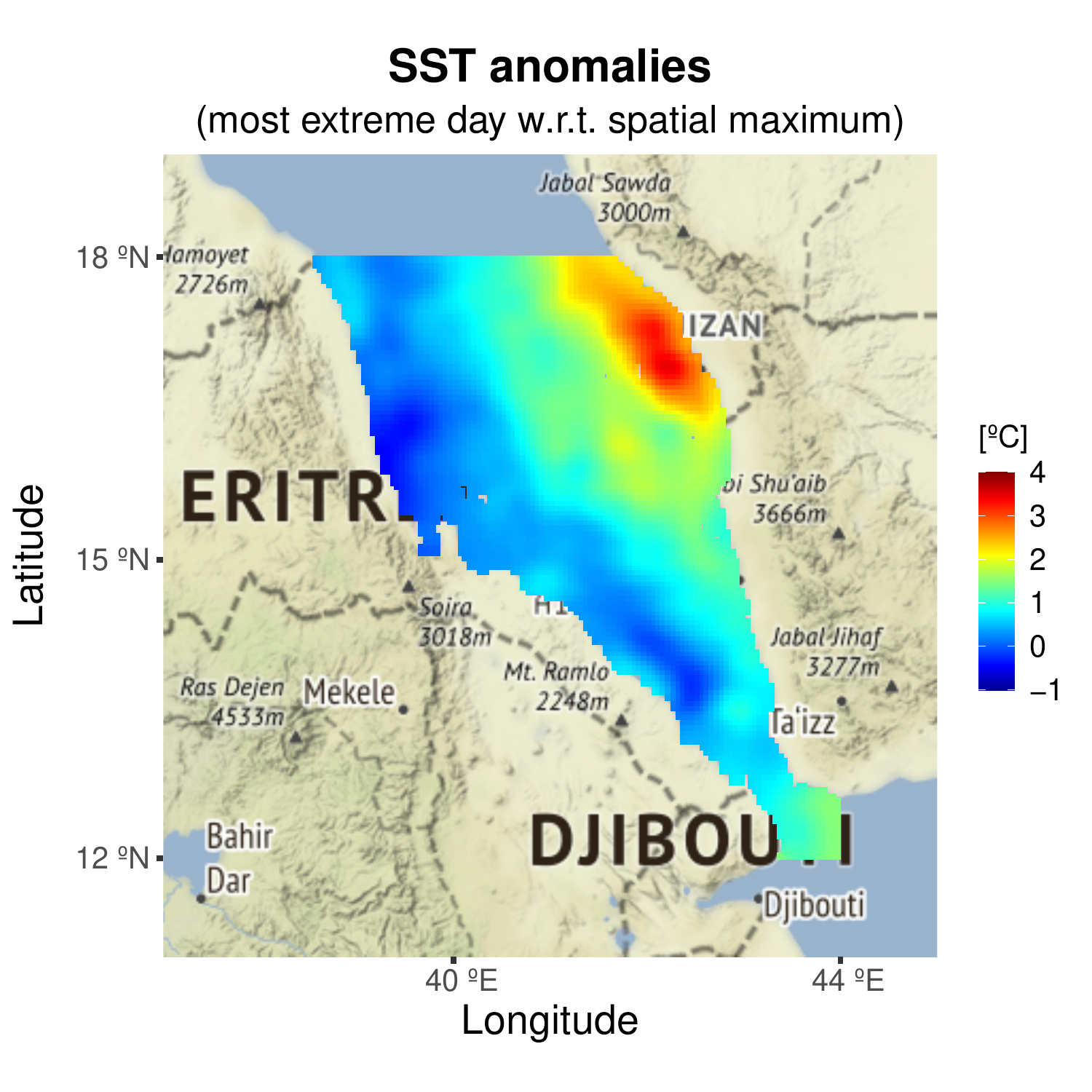} 
	\caption{The study region in the Southern Red Sea consisting of 5562 grid cells. The plots show the sea surface temperature anomalies (i.e., the detrended SST data) for the most extreme day with respect to the spatial median (left) and maximum (right), respectively.} \label{fig:map}
\end{figure}

In order to demonstrate the behavior of our estimators for finite sample sizes, we apply them to a simulated realization of a regularly varying space-time stochastic process. Here, we choose a stationary space-time Brown--Resnick process \citep{brown-resnick-77,kabluchko-etal-09,davis-etal-13,huser-davison-14,engelke-etal-15} with unit Fr\'echet margins, simulated on a grid $S$ consisting of $|S|=1\,367$ grid points in an irregular domain delimited by a box with coordinates in $40.0$--$42.5^\circ$E and $16.5$--$18.0^\circ$N. That is, the spatial domain is a subset the southern Red Sea region mapped in Figure~\ref{fig:map} and considered in the application in Section~\ref{sec:application}, while the length of the time series is $12\,000$, also similar to the application. The properties of the Brown--Resnick process are determined by the variogram of the underlying space-time Gaussian process, which we choose here to be separable with
	      $$ \gamma(\bm h,t) = \left\| \begin{pmatrix}2.6 h_1 \\ 2.4 h_2\end{pmatrix} \right\|^{1.9} + |t|^{1.1}, \qquad \bm h=\begin{pmatrix}h_1 \\ h_2\end{pmatrix}, \bm s\in S, \ t=1,\ldots,12\,000.$$
The variogram parameters were chosen so that the dependence structure of the resulting Brown--Resnick process resembles that of the sea temperature data in the data in Section~\ref{sec:application}.

The simulation makes use of the extremal functions algorithm of \cite{dombry-etal-16}. Here, note that an exact simulation of the max-stable process would on average require the simulation of $|\mathcal{S}| \times |\{1,\ldots,12\,000\} = 1\,367 \times 12\,000$ Gaussian processes on the whole spatio-temporal domain $\mathcal{S} \times \{1,\ldots,12\,000\}$ and, thus, would be extremely time-consuming. Therefore, we use an approximate version of the exact algorithm of \cite{dombry-etal-16}, which ensures exactness only on a subgrid consisting of every second point, similarly to \cite{oesting-strokorb-22}. Furthermore, we assume that an extremal function cannot have an influence over a temporal lag exceeding $18$, that is, the extremal function at the space-time point $(\bm s,t)$ is simulated on a smaller domain $\mathcal{S} \times \{t-18,\ldots,t-1,t,t+1,\ldots,t+18\}$ only. Thus, the computing time for the simulation is reduced significantly, while the accuracy is hardly affected, as preliminary experiments on smaller domains indicate.

As the simulated process is regularly varying with (heavy-tailed) Fr\'echet margins, the theory developed above applies and allows us to use any homogeneous functional $r$ for defining extremes. Here, we illustrate our methodology with the spatial average
$$ r(f) = \frac{1}{|S|} \sum\nolimits_{\bm s \in S} f(s), $$
as the corresponding tail process for space-time Brown--Resnick processes turns out to be easy to simulate from \citep[see][for instance]{dombry-etal-16}. 

\begin{figure}[t!]
\centering
\includegraphics[width=0.8\textwidth]{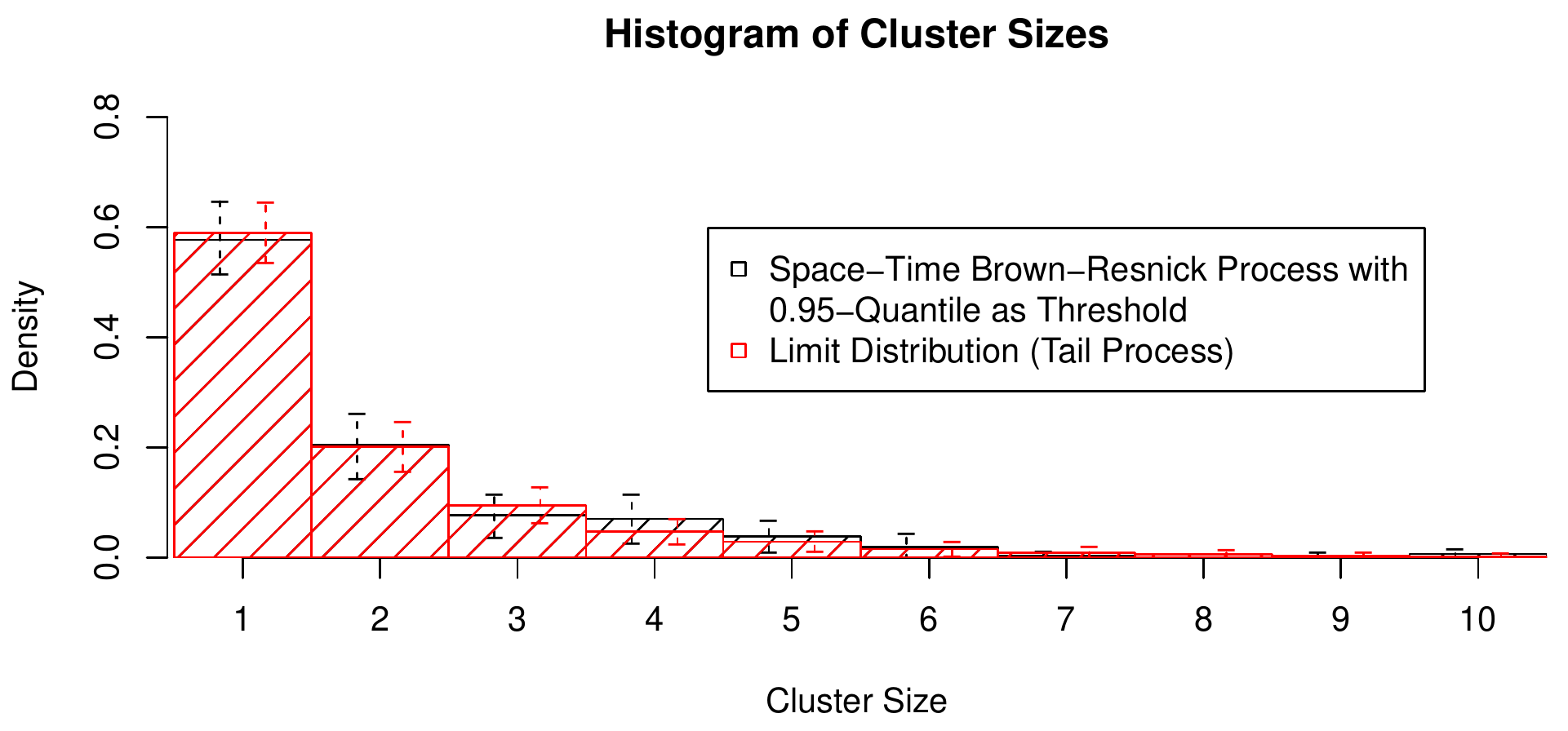}
\caption{Cluster size distribution for the simulated example. The black bars correspond to the estimated probabilities with the 95\% confidence intervals obtained by the multiplier block bootstrap, while the red bars are the result of Monte Carlo simulations from the tail process with the corresponding confidence intervals being calculated according to Prop.~\ref{prop:asympt-normal}.} \label{fig:simu_clustersize}
\end{figure}

\begin{figure}[t!]
	\centering
        \includegraphics[width=0.8\textwidth]{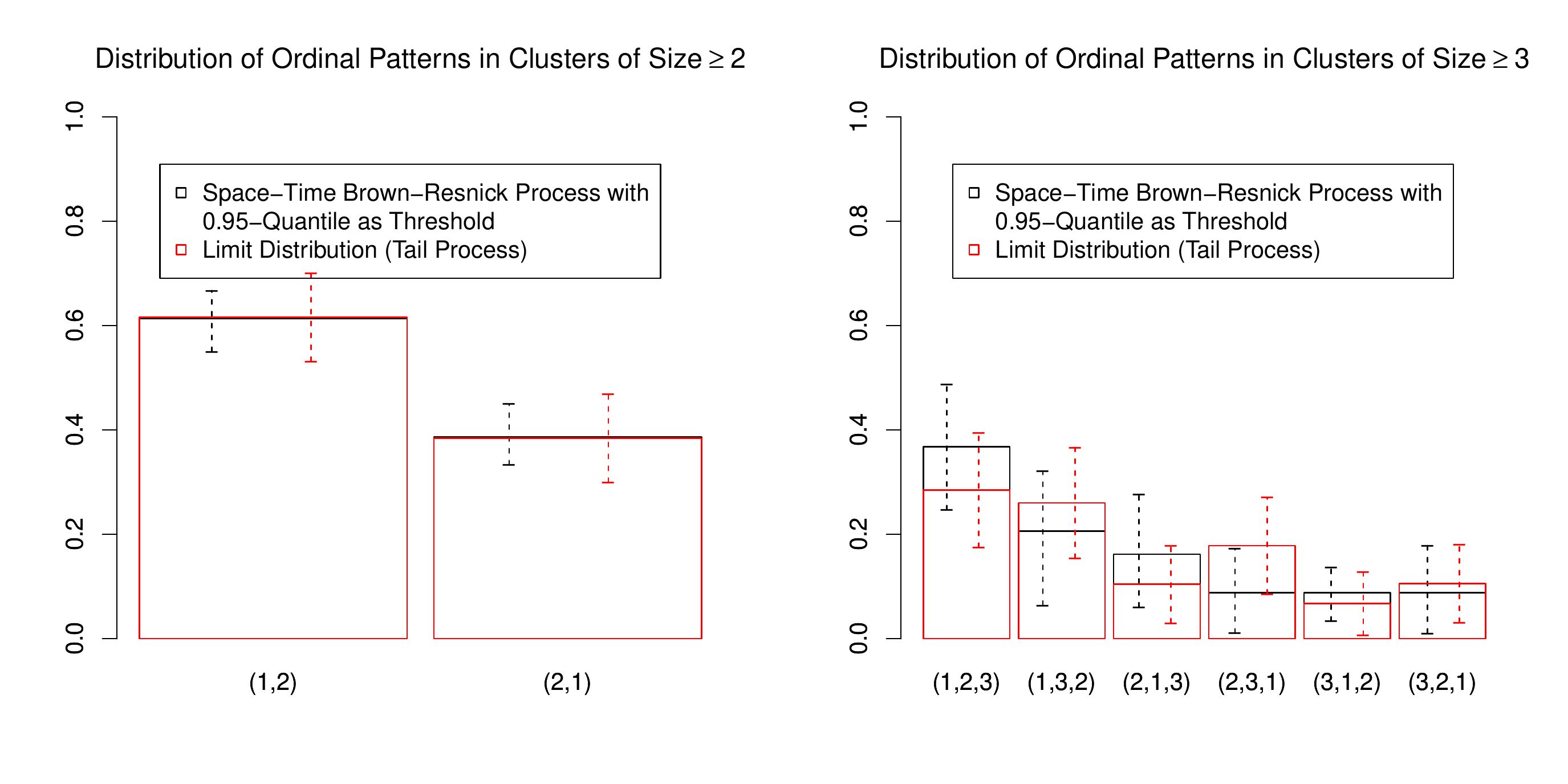}
	\caption{Distribution of the ordinal pattern for the beginning of clusters of size $\ell \geq 2$ (left) and $\ell \geq 3$ (right), for the simulated examples. Analogously to Figure~\ref{fig:simu_clustersize}, the black bars correspond to the estimated probabilities with 95\% confidence intervals obtained via bootstrap and the red bars are the result of Monte Carlo simulations from the tail process with confidence intervals according to Prop.~\ref{prop:asympt-normal}.}  \label{fig:simu_patterns}
\end{figure}

We apply our proposed ratio estimators to estimate the following two quantities:
\begin{itemize}
    \item the cluster size distribution, i.e., $\PP(C_u^r=\ell)$, $\ell=1,2,\ldots$ (recall Section~\ref{sec:clustersize}),
    \item the distribution of ordinal patterns with respect to the actual risk functional $r$ (recall Section~\ref{sec:lim-riskfun}); here, we focus on the patterns of the first $\ell$ time points within a cluster, i.e., we focus on all clusters of size greater than or equal to $\ell$, with $\ell=2,3$,
\end{itemize}
based on the choice of the threshold $u$ as the $95$th percentile of $r(\bm X_t)$, $t=1,\ldots,12\,000$. For each estimate, the corresponding $95\%$ confidence interval is obtained using the multiplier block bootstrap described at the end of Section~\ref{sec:estim} with a block length of $1\,000$. The results are displayed in Figure~\ref{fig:simu_clustersize} and Figure~\ref{fig:simu_patterns}, respectively.

Our results based on estimators of the form $\widehat R_{n,u_n}(A,A_0)$ can then be compared to the theoretical limit $\mu_{\{-1,\ldots,\ell\}}(A)/\mu_{\{-1,\ldots,\ell\}}(A_0)$ which can be accurately approximated via direct Monte Carlo simulations from the tail process. Similarly, the asymptotic variance in Prop.~\ref{prop:asympt-normal} can also be calculated precisely. Assuming asymptotic normality, theoretical confidence intervals can thus be derived, as well.

From Figures~\ref{fig:simu_clustersize} and \ref{fig:simu_patterns}, it can be seen that the estimated distribution of the cluster size and the ordinal patterns in clusters of size at least $2$ are very close to the theoretical values. For patterns of size at least $3$, there are somewhat larger deviations, which can be explained by the fact that even in a time series of length $12\,000$, the number of long clusters is still very small. As can be seen, the bootstrap-based confidence intervals are fairly wide, but still cover the theoretical values, which indicates that our estimator performs well overall. Comparing the width of the asymptotic confidence intervals and the bootstrap-based ones, there are slight differences, but no systematic ones. This suggests that the bootstrap procedure also works satisfactorily.

\section{Application to Red Sea surface temperature data} \label{sec:application}

Sea surface temperature (SST) is known to be a major factor influencing the survival of marine life and the sustainability of fragile ecosystems, such as coral reefs  \citep{reaser2000coral,berumen2013status,lewandowska2014effects}. Therefore, the spatiotemporal analysis of extreme SST data is important, and an entire special issue of the journal \emph{Extremes} was devoted to the spatiotemporal prediction of extremes with application to SST data over the whole Red Sea; see \url{https://link.springer.com/journal/10687/volumes-and-issues/24-1} and the Editorial by \citet{Huser:2021}. Apart from spatiotemporal prediction, previous research has also focused on identifying time trends in SST data due to climate change and estimating extreme SST hotspots \citep{Hazra.Huser:2021}, as well as on the modeling of spatial and spatiotemporal extremal dependence in SST data \citep{Simpson.etal:2022,huser-etal-22}. However, the characterization of complex spatiotemporal patterns and the within-cluster behavior in functional time series of SST extremes has never been investigated. We here re-analyze the Red Sea data used in the above papers to illustrate our proposed methodology. Our analysis of spatiotemporal patterns complements previous studies by providing new insights into the complex hydrodynamic behavior of the SST data across the study domain, and sheding light on the spatial extent and persistence of extreme SST events. 

The dataset we use comes from the Operational Sea Surface Temperature and Sea Ice Analysis (OSTIA) project, producing satellite-derived daily SST data at $0.05^\circ \times 0.05^\circ$ resolution \citep{donlon2012operational}. Over the whole Red Sea, daily SST data are available at 16703 grid cells between 1985--2015 (i.e., 31 years worth of daily data) and we here focus on the southern Red Sea, below latitude $18^\circ$N, resulting in 5562 high-resolution grid cells in total. Figure~\ref{fig:map} shows the study region and SST data for two days that are considered as extreme with respect to different functionals.
	
Our methodological framework assumes temporal stationarity, so we first need to appropriately detrend the data. Let $\{\mathcal Y(\bm s,t)\}_{\bm s\in S,t\in T}$ denote the sea surface temperature process over the study domain $S$ (comprising 5562 pixels) for the time period $T=\{1,\ldots,11315\}$. We assume that the data can be described, at each location $\bm s$, by the linear regression model
$$\mathcal Y(\bm s,t)=\beta_0(\bm s)+\beta_1(\bm s)\,t + \sum\nolimits_{j=1}^{12}\beta_{1+j}(\bm s)\psi_j(t)+X(\bm s,t),\qquad t=1,\ldots,11315,$$
where $\{\psi_j(t)\}$ denote 12 cyclic cubic splines (i.e., one for each month) capturing the seasonal cycle in a flexible way, $t$ is a linear time trend capturing the effect of climate change, $\{\beta_j(\bm s)\}_{j=0,\ldots,13}$ are site-specific regression coefficients to estimate from the data, and the process $\{X(\bm s,t)\}_{\bm s\in S,t\in T}$ is a zero-mean stochastic residual component, assumed to be stationary in time, but not necessarily in space. We estimate regression parameters by least squares at each site separately, pooling information from neighboring grid cells within a disk of radius $30$\,km. This local inference approach gives high flexibility to estimate complex spatiotemporal trends, while the pooling of information allows one to smooth estimates spatially and borrow strength across nearby sites to reduce the overall uncertainty. 

We then apply our new spatiotemporal pattern estimation approach to $\{\hat x(\bm s,t)\}_{s\in S,t\in T}$, pseudo-observations from the residual component $\{X(\bm s,t)\}_{s\in S,t\in T}$ that are obtained by subtracting the estimated SST mean from the realizations $\{y(\bm s,t)\}_{s\in S,t\in T}$ of $\{\mathcal Y(\bm s,t)\}_{s\in S,t\in T}$ (after replacing the $\beta_j(\bm s)$ coefficients with their least square estimates). Note that the residuals $\hat x(\bm s,t)$ can be interpreted as the SST anomaly, and extremes from $\hat x(\bm s,t)$ represent abnormal sea temperature excesses with respect to the usual SST distribution at the specific space-time point $(\bm s,t)$. We also considered rescaling $\hat x(\bm s,t)$ by its standard deviation (which varies moderately over space and time), but this makes the interpretation of the results more difficult and we thus proceed with $\hat x(\bm s,t)$ (detrended, but not rescaled).

Remind that, even though $X$ may not be regularly varying itself because SST data often have a bounded upper tail rather than a heavy tail, our methodology can still be applied for a variety of functionals, as long as some marginal transformation of $X$ is regularly varying. This condition is equivalent to $X$ being in the max-domain of attraction of some max-stable process, as explained throughout Sections~\ref{sec:risk-fun-regvar} and \ref{sec:inside-cluster}. In some applications, this may be considered quite restrictive as it implies asymptotic dependence. However, extremes in sea surface processes are usually strongly spatially dependent, even at large distances, and they also tend to persist over consecutive days, if not weeks. Therefore, asymptotic dependence is a natural assumption here, and statistical models with this property have in fact already been applied to this dataset \citep[see, e.g.,][]{Hazra.Huser:2021,huser-etal-22}.

\begin{figure}[t!]
        \centering
	\includegraphics[width=0.8\textwidth]{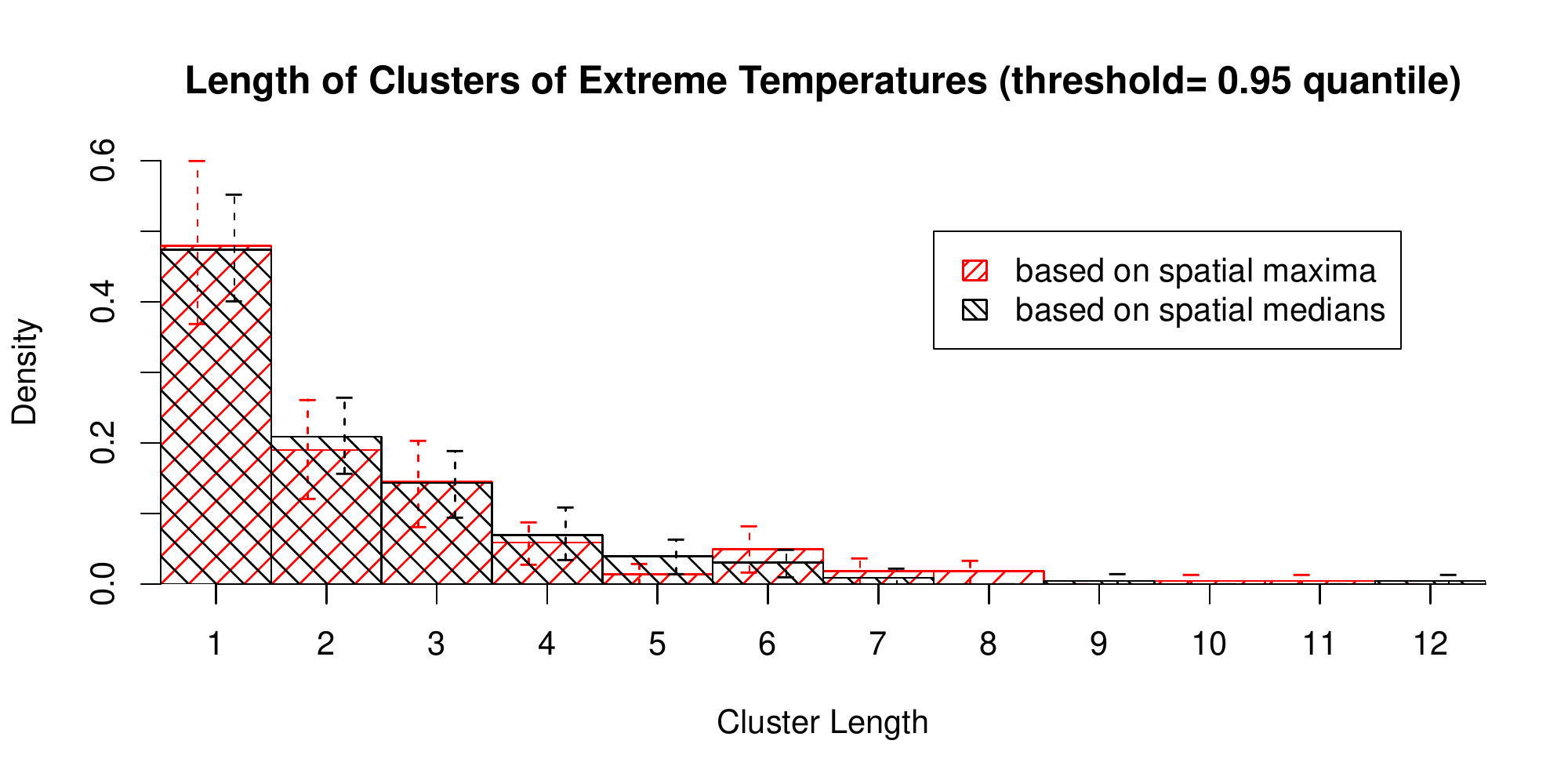} 
	\caption{Cluster size distribution for extreme SST defined as high spatial maxima (red) and spatial medians (black) exceeding the 95\% empirical quantile. Vertical dashed segments display $95\%$ confidence intervals, obtained via our proposed multiplier block bootstrap.} \label{fig:clustersize}
\end{figure}

We here choose to consider two different risk functionals, namely
\begin{itemize}
\item the spatial maximum, $r_{\rm max}(f) = \max_{\bm s \in S} f(\bm s)$, and 
\item the spatial median, $r_{\rm med}(f) = {\rm median}_{\bm s \in S} f(\bm s)$.
\end{itemize}
At first, we estimate the length, $\ell$, of extremal clusters defined as consecutive spatial fields $f_0,\ldots,f_{\ell-1}$, such that $r_{\rm max}(f_{-1})\leq u,r_{\rm max}(f_i)>u,r_{\rm max}(f_\ell)\leq u$,
 for all $i=0,\ldots,\ell-1$, or $r_{\rm med}(f_{-1})\leq u,r_{\rm med}(f_i)>u,r_{\rm med}(f_\ell)\leq u$, for all $i=0,\ldots,\ell-1$, where $u$ is a high threshold, here taken as the empirical 95\% quantile of the data $\{\hat x(\bm s,t)\}_{\bm s\in S,t\in T}$. Figure~\ref{fig:clustersize} reports the estimated cluster size distribution. The results for the risk functional defined as the spatial maximum and those for the spatial median are very similar to each other: about $35\%$--$40\%$ of clusters comprise individual events, $20\%$--$25\%$ are of size two, $15\%$--$20\%$ of size three, and they are very rarely longer than a week (7 consecutive days).

\begin{figure}[t!]
        \centering
	\includegraphics[width=0.8\textwidth]{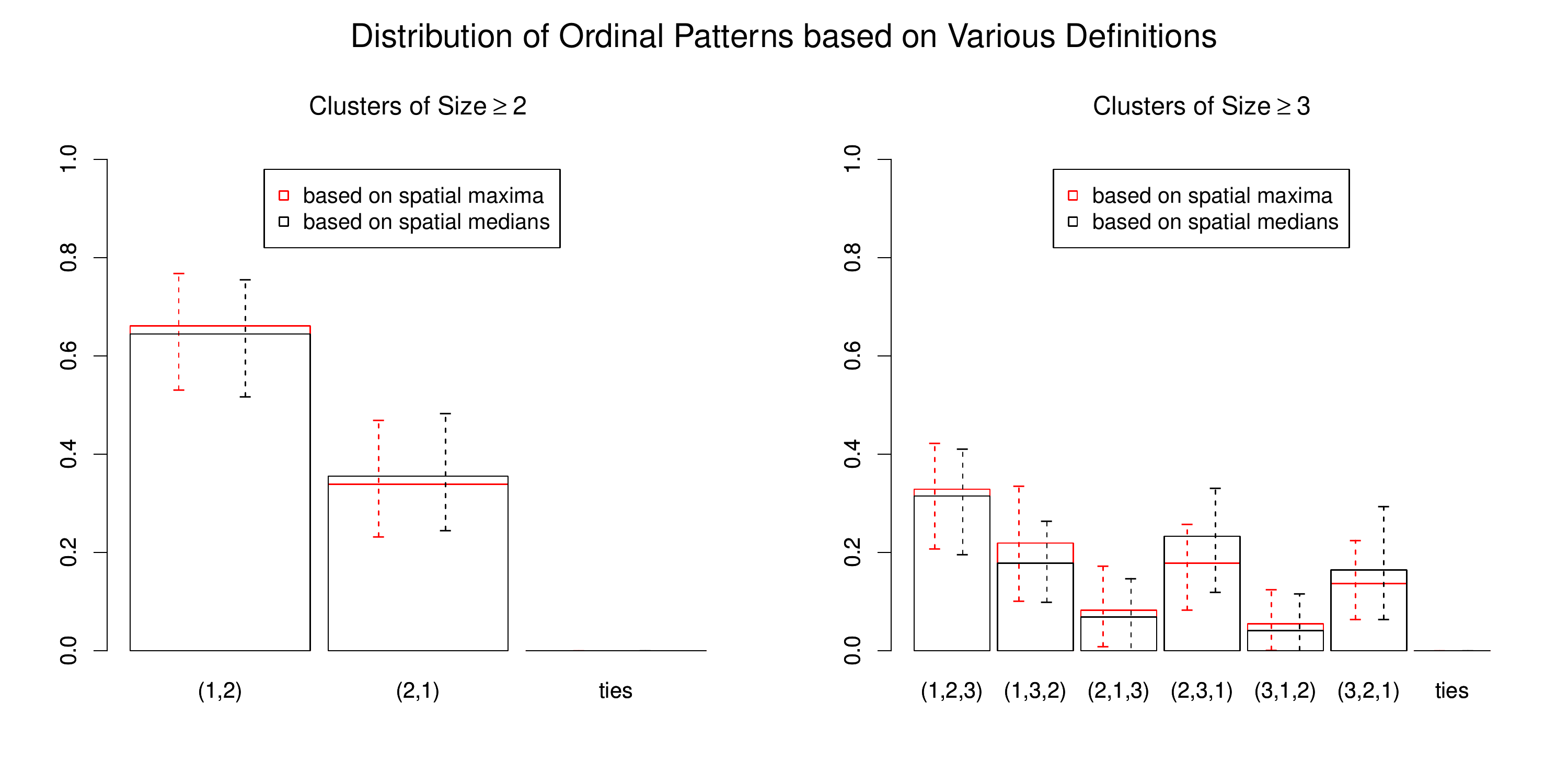} 
	\caption{Bar charts displaying the distribution of ordinal patterns with respect to the intensity of extreme SST defined through the spatial maximum (red) and spatial median (black), for the beginning of clusters of size at least $2$ (left) and $3$ (right). Vertical dashed segments display $95\%$ confidence intervals, obtained via our proposed multiplier block bootstrap.} \label{fig:patterns}
\end{figure}

We then zoom into the clusters themselves and estimate ordinal patterns defined in terms of the respective risk functionals, giving an indication of how the intensity of the extreme event evolves through time. Figure~\ref{fig:patterns} displays, for each risk functional, the distribution of these ordinal patterns for the beginning of clusters of size at least two, and at least three. While we could restrict ourselves to clusters of size \emph{exactly} two or three, the resulting sample size would be too low for accurate estimation. Again, there is strong similarity between results for the spatial maximum and spatial median. The ordinal pattern $(1,2)$ appears about $60\%$--$70\%$ of the time, indicating that the value of the risk functional at the start of a cluster of size at least two tends to increase. In other words, spatial extreme events that persist over consecutive days tend to strengthen at the beginning, which makes sense from a physical perspective. A similar story holds for clusters of size at least three, with the increasing ordinal pattern $(1,2,3)$ occurring more than 30\% of the time. Interestingly, the ordinal patterns $(2,1,3)$ and $(3,1,2)$, which describe a ``V-shape'' (i.e., with a decrease followed by an increase) are the least likely. Conversely, the patterns $(1,3,2)$ and $(2,3,1)$ describing an ``inverse V-shape'' characteristic of relatively short clusters are the most likely combined. While these results make sense, our methodology is helpful to precisely quantify the frequency of these complex within-cluster behaviors.

\begin{figure}[t!]
    \centering
    \includegraphics[width=0.8\textwidth]{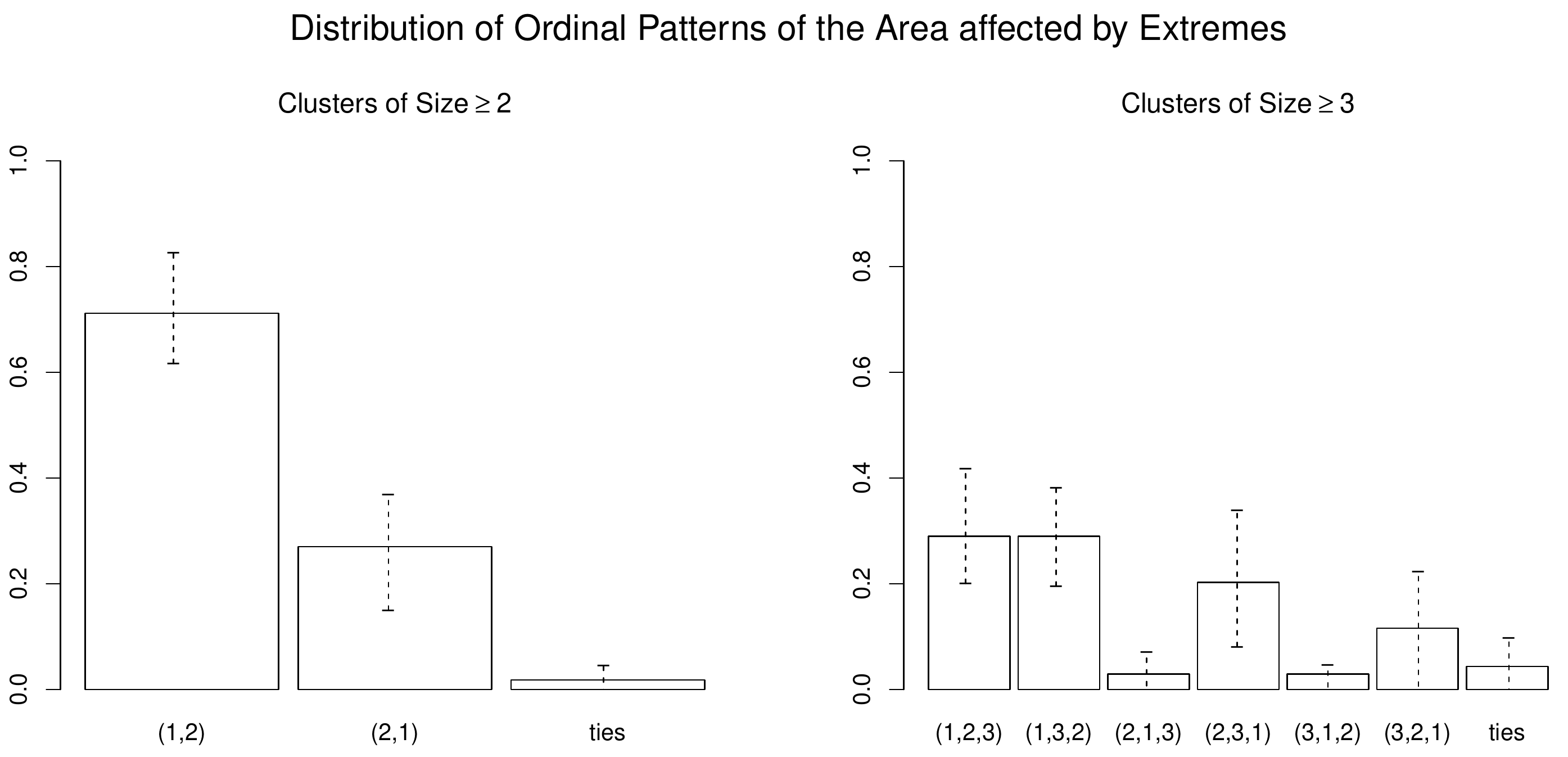}  \\
	\caption{Bar charts displaying the distribution of ordinal patterns with respect to the relative area affected by extreme SST, based on exceedances of the spatial maximum for the beginning of clusters of size at least $2$ (left) and $3$ (right). Vertical dashed segments display $95\%$ confidence intervals, obtained via our proposed multiplier block bootstrap.} \label{fig:patterns-area}
\end{figure}

To investigate further within-cluster characteristics, we then estimate the distribution of patterns for the relative area affected by extremes, i.e., the number of grid cells within a spatial field exceeding the chosen threshold divided by the total number of grid cells. We here only consider clusters defined by the spatial maximum. Note that the estimated area affected by extremes cannot be zero, given that we consider clusters of exceedances (i.e., the maximum exceeds the threshold). Figure~\ref{fig:patterns-area} reports the results. The results are similar to those discussed above, with the increasing pattern $(1,2)$ being the most frequent one. This indicates that the spatial extent of extreme events also tends to increase at the beginning of a cluster of size at least two. Similar results hold for clusters of size at least three.

\begin{figure}[t!]
    \centering
    \includegraphics[width=0.8\textwidth]{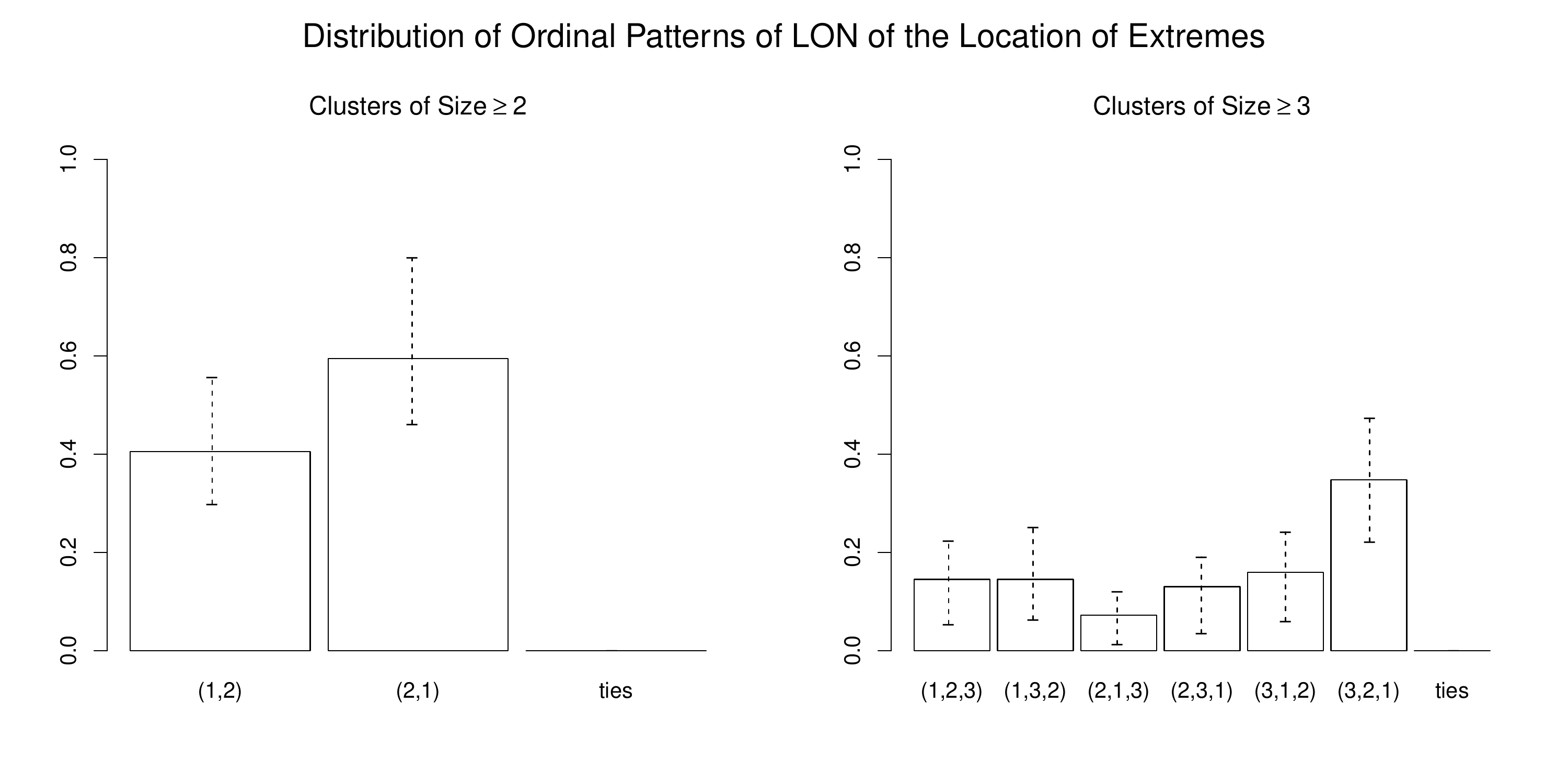}
    \includegraphics[width=0.8\textwidth]{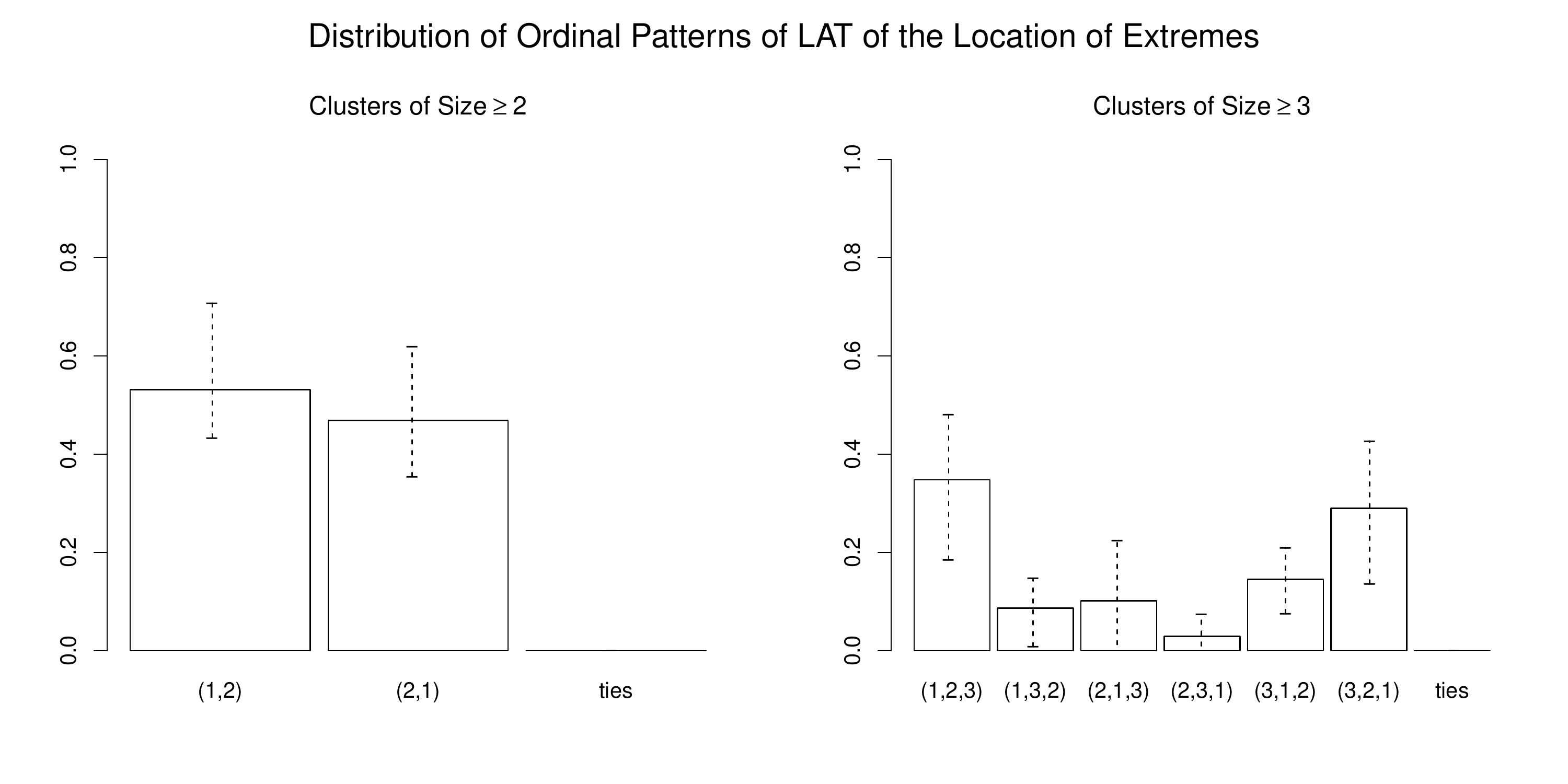}
	\caption{Bar charts displaying the distribution of ordinal patterns with respect to the longitude (top) and latitude (bottom) of the centroid of the region of the most extreme temperatures exceeding the overall $95\,\%$ quantile for the beginning of clusters of extreme SST of size at least $2$ (left) and $3$ (right). Vertical dashed segments display $95\%$ confidence intervals, obtained via our proposed multiplier block bootstrap.} \label{fig:patterns-location}
\end{figure}

Finally, to investigate the movement of high sea temperatures across the study region during a spatial extreme episode, we estimate the distribution of ordinal patterns for the latitude and longitude (Figure~\ref{fig:patterns-location}) of the centroid of the area affected by extremes (as defined in Section~\ref{sec:lim-loc}). The results indicate that when a spatiotemporal extreme event occurs, the area of highest SST intensity tends to move from South to North, and from East to West, most of the time. These results seem to be broadly aligned with the southern Red Sea circulation literature and might reflect the formation of eddies \citep{raitsos-etal-13,zhan-etal-14}, though further in-depth analyses would be required to confirm this. In future research it would be interesting to exploit our methodology to study these hydrodynamic properties in more detail using alternative specifically customized functionals.

\section{Conclusion} \label{sec:conclusion}
In order to investigate the probabilistic behavior of temporal clusters formed by spatio-temporal extreme events, and to understand the complex within-cluster behavior of the process under study with respect to various spatial summaries of interest, we have developed a novel non-parametric method that allows accurately estimating the limiting distributions of such quantities in a flexible and computationally-efficient way. 

Our proposed methodology relies on the theoretical framework of functional regular variation for reliable joint tail extrapolation beyond the range of the data, and it is based on the notion of risk functionals to account for various definitions of extreme events. While our work has mostly focused on describing the within-cluster behavior of extreme events (i.e., their temporal evolution) in terms of the ordinal patterns of various functional characteristics, our theoretical results apply more generally. The characteristics that our framework can track over time are very diverse, including the magnitude of the extreme event (characterized by the risk functional itself), spatial risk measures (e.g., the size of the affected area), and the location of the extreme event (characterized by various types of location measures). Our framework is, however, not limited to these characteristics. In future research, it would be interesting to extend our method to estimate the within-cluster behavior of an extreme event with respect to \emph{spatio-temporal} functionals of the process capturing, for example, its local rotational circulation properties. This could prove helpful for identifying the fingerprints of well-known physical phenomena, such as circular currents of water (i.e., eddies) in data analyses similar to our sea surface temperature application. 

In contrast to state-of-the-art parametric extreme-value models, the great benefit of our non-parametric approach is that it is not subject to stringent assumptions (e.g., spatial stationarity and isotropy, autoregressive temporal dependence structure, etc.) and can thus be used to flexibly estimate complex spatio-temporal characteristics in a wide range of real data applications. Moreover, since our proposed estimators have a very simple form, they can be computed efficiently, even in ultra-high dimensions. For example, in our Red Sea surface temperature application, the full dataset comprises about 63 million dependent spatio-temporal variables, and our results can be obtained in just a few seconds. Furthermore, our proposed approach is not only feasible in such high dimensions, but it also strongly benefits from massive datasets, as our non-parametric estimators then become more accurate.

Although regular variation is a key assumption here, our results still hold under monotone marginal transformations of the process under study (e.g., when it is expressed on a different marginal scale) for a wide range of functionals (i.e., specifically, those that are invariant under monotone transformations). This implies that our methodology is not strictly restricted to marginally heavy-tailed processes, but it can also be applied more generally to spatio-temporal processes exhibiting asymptotic dependence with light or bounded marginal tails. The asymptotic dependence assumption, however, is a crucial condition that we cannot easily bypass. In our sea surface temperature data application, this does not seem to be a limitation, but empirical evidence has shown that some other environmental processes tend to support asymptotic independence instead. In future research, it would therefore be interesting to extend our methodology to the framework of hidden regular variation, in order to accommodate both asymptotic dependence and independence in a flexible way.

\section*{Supplementary Materials}

The supplementary materials include the proofs of Propositions \ref{prop:risk-fun}, \ref{prop:spat-risk} and \ref{prop:location}. 

\section*{Funding}

MO gratefully acknowledges partial support by the project “Climate Change and Extreme Events - ClimXtreme Module B - Statistics (subproject B3.1)” funded by the German Federal Ministry of Education and Research (BMBF) with the grant number 01LP1902I.
RH was partially supported by funding from King Abdullah University of Science and Technology (KAUST) Office of Sponsored Research (OSR) under Award No.~OSR-CRG2020-4394.

\bibliographystyle{abbrvnat}
\bibliography{lit}

\newpage

\appendix

\section*{Supplementary Material}

\subsection*{Proof of Proposition \ref{prop:risk-fun}}

  Due to the continuity of $r$, there exists some constant $c_r > 0$ such
  that $r(f) \leq c_r \cdot \|f\|_\infty$ for all $f \in C_+(S)$. Consequently,
  $r(\bm X_0) > x$ implies that $\|\bm X_0^*\|_\infty > c_r^{-1}x$. Thus, for any
  $a > 1$ and any measurable set $A \subset C_+(S)^\ZZ$, we obtain
  \begin{align*}
   & \PP\left(r(\bm X_0) > ax, \, 
        \left\{ \frac{r(\bm X_t)}{r(\bm X_0)}\right\}_{t \in \ZZ} \in A
        \, \Big| \, r(\bm X_0) > x \right)\\
   ={}& \frac{\PP\left(r(\bm X_0) > ax, \,
   	\left\{ \frac{r(\bm X_t)}{r(\bm X_0)}\right\}_{t \in \ZZ} \in A
   	\, \Big| \, \|\bm X_0\|_\infty > c_r^{-1}x \right)}
    {\PP\left(r(\bm X_0) > x
    	\, \Big| \, \|\bm X_0\|_\infty > c_r^{-1}x \right)} 
   {}\to{} \frac{\PP\left(P r(\bm \Theta_0) > c_r a, \,
   	\left\{ \frac{r(\bm \Theta_t)}{r(\bm \Theta_0)}\right\}_{t \in \ZZ} \in A \right)}
   {\PP\left(P r(\bm \Theta_0) > c_r\right)} \\
   ={}& \EE\left( \left[\frac{r(\bm \Theta_0)}{c_r}\right]^\alpha \right)^{-1}
        \cdot 
        \EE\left( \left[\frac{r(\bm \Theta_0)}{c_r a}\right]^\alpha \cdot
          \mathbf{1}\left\{ \left\{ \frac{r(\bm \Theta_t)}{r(\bm \Theta_0)}\right\}_{t \in \ZZ} \in A \right\} \right)
   {}={} a^{-\alpha} \cdot \PP(\{\bm \Theta_t^{r}\}_{t\in\ZZ} \in A),   
  \end{align*} 
  as $x \to \infty$, where we used that $r(\bm \Theta_t) \leq c_r$ a.s.\ for all
  $t \in \ZZ$.
\qed

\subsection*{Proof of Proposition \ref{prop:spat-risk}}
Using the same constant	$c_r > 0$ as in the proof of Prop.~\ref{prop:risk-fun},
we obtain
\begin{align*}	
 & \PP\big( 
 (m(\{\bm X_0>u\}),\ldots,m(\{\bm X_{\ell-1}>u\})) \in A 
 \mid \\
 & \qquad \qquad r(\bm X_{-1}) \leq u, r(\bm X_0) > u, \ldots, r(\bm X_{\ell-1}) > u, r(\bm X_\ell) \leq u\big) \\
 ={}& [\PP\left(r(\bm X_{-1}) \leq u, r(\bm X_0) > u, \ldots, r(\bm X_{\ell-1}) > u,
 	                    r(\bm X_\ell) \leq u \mid \|\bm X_0\| > c_r^{-1}u \right)]^{-1}\\
 	& \quad \cdot \PP\big( (m(\{\bm X_0>u\}),\ldots,m(\{\bm X_{\ell-1}>u\})) \in A, \\
 	& \qquad \qquad r(\bm X_{-1}) \leq u, r(\bm X_0) > u, \ldots, r(\bm X_{l-1}) > u,
 	                r(\bm X_\ell) \leq u  \mid \|\bm X_0\| > c_r^{-1}u \big) \\
  \stackrel{u \to \infty}{\longrightarrow}{}&
  [\PP\left( P r(\bm \Theta_{-1}) \leq c_r,
  P \cdot r(\bm \Theta_0) > c_r, \ldots, P r(\bm \Theta_{\ell-1}) > c_r,
   P \cdot r(\bm \Theta_\ell) \leq c_r\right)]^{-1}\\
  & \quad \cdot \PP\big( 
 	(m(\{P \bm \Theta_0 > c_r\}), \ldots, 
 	 m(\{P \bm \Theta_t > c_r\})) \in A, \\
   &  \qquad \qquad P r(\bm \Theta_{-1}) \leq c_r,
 	  P \cdot r(\bm \Theta_0) > c_r, \ldots, P r(\bm \Theta_{\ell-1}) > c_r,
 	  P \cdot r(\bm \Theta_\ell) \leq c_r\big).
\end{align*}
While the first terms equals $c_r^\alpha/\EE[\min\{1, r(\bm \Theta_{1})^\alpha, \ldots, r(\bm \Theta_{\ell-1})^\alpha\} 
- \max\{r(\bm \Theta_{-1})^\alpha,r(\bm \Theta_\ell)^\alpha\}]_+$, the second term can be 
rewritten as
\begin{align*}
 & \int_1^\infty
 \PP\big( (m(\{\bm \Theta_0 > c_r/y\}),\ldots,m(\{\bm \Theta_{\ell-1} > c_r/y\})) \in A,\\
 & \qquad \qquad \quad
      r(\bm \Theta_{-1}) \leq c_r/y, r(\bm \Theta_0) > c_r/y, \ldots, r(\bm \Theta_{l-1}) > c_r/y, r(\bm \Theta_l) \leq c_r/y \big) \cdot
      \alpha y^{-\alpha-1} \, \mathrm{d}y \\
 ={}& c_r^{-\alpha} \int_0^{c_r^\alpha}
 \PP\big( (m(\{\bm \Theta_0^\alpha > \eta\}), \ldots, m\{\bm \Theta_t^\alpha > \eta\}) \in A,\\
 & \qquad \qquad \qquad r(\bm \Theta_{-1})^\alpha \leq \eta, r(\bm \Theta_0)^\alpha > \eta, \ldots, r(\bm \Theta_{\ell-1})^\alpha > \eta, r(\bm \Theta_\ell)^\alpha \leq \eta \big)
 \, \mathrm{d}\eta   
\end{align*}
where we substituted $\eta= (c_r/y)^\alpha$. On the one hand, $\one\{\bm \Theta_0^\alpha > \eta\} = \bm 0_S$ a.s.\ for all $\eta>1$ and, consequently, 
$(m(\{\bm \Theta_0^\alpha > \eta\}), \ldots, m\{\bm \Theta_{\ell-1}^\alpha > \eta\}) \notin A \subset (0,\infty)^\ell$. On the other hand, $r(\bm \Theta_0)^\alpha \leq c_r^\alpha$ with probability one. Thus, 
\begin{align*}
& \int_0^{c_r^\alpha}
\PP\big( (m(\{\bm \Theta_0^\alpha > \eta\}), \ldots, m(\{\bm \Theta_t^\alpha > \eta\})) \in A,\\
& \qquad \qquad \qquad r(\bm \Theta_{-1})^\alpha \leq \eta, r(\bm \Theta_0)^\alpha > \eta, \ldots, r(\bm \Theta_{\ell-1})^\alpha > \eta, r(\bm \Theta_\ell)^\alpha \leq \eta \big)
\, \mathrm{d}\eta \\
={}& \int_0^1
\PP\big( (m(\{\bm \Theta_0^\alpha > \eta\}), \ldots, m(\{\bm \Theta_{\ell-1}^\alpha > \eta\})) \in A,\\
& \qquad \qquad \qquad r(\bm \Theta_{-1})^\alpha \leq \eta, r(\bm \Theta_0)^\alpha > \eta, \ldots, r(\bm \Theta_{\ell-1})^\alpha > \eta, r(\bm \Theta_\ell)^\alpha \leq \eta \big)
\, \mathrm{d}\eta. 
\end{align*}
Furthermore, we note that the above conditions on the 
joint distribution of the random vector $m(\{\bm \Theta_0>\eta\}), \ldots,
m(\{\bm \Theta_{\ell-1}>\eta\})$ for $\eta \in [0,1]$ ensure the existence of the integral
as well as the convergence of the expression above.
\qed

\subsection*{Proof of Proposition \ref{prop:location}}

 The proof runs analogously to the proof of 
 Prop.~\ref{prop:spat-risk}.
 Using the same constant $c_r > 0$ from Prop.~\ref{prop:risk-fun}, we obtain
 \begin{align*}
  \PP( (c&(u^{-1}\bm X_0), \ldots, c(u^{-1}\bm X_{\ell-1})) \in A \mid 
   r(\bm X_{-1}) \leq u, r(\bm X_0) > u, \ldots, r(\bm X_{\ell-1}) > u, r(\bm X_\ell) \leq u)\\ 
 ={}&  \left[\PP\left(r(\bm X_{-1}) \leq u, r(\bm X_0)>u, \ldots, r(\bm X_{\ell-1})>u,
 	       r(\bm X_\ell) \leq u \, \Big| \, \|\bm X_0\|_\infty > c_r^{-1}u \right)\right]^{-1} \\
   & \cdot \PP\big( (c(u^{-1}\bm X_0), \ldots, c(u^{-1}\bm X_{\ell-1})) \in A, \\
 	& \qquad \qquad r(\bm X_{-1}) \leq u, r(\bm X_0) > u, \ldots, r(\bm X_{\ell-1}) > u, r(\bm X_\ell) \leq u
 	\, \big| \, \|\bm X_0\|_\infty > c_r^{-1}u \big) \\
 \stackrel{u \to \infty}{\longrightarrow}{}&
  \left[\PP(P \max\{1, r(\bm \Theta_1), \ldots, r(\bm \Theta_{\ell-1})\} \leq c_r,
        	 P \min\{r(\bm \Theta_{-1}), r(\bm \Theta_\ell)\} > c_r)
  \right]^{-1} \\
  & \cdot 
 \PP\big((\widetilde c(\bm \Theta_0, \one\{P \bm \Theta_0 > c_r\}), \ldots, \widetilde c(\bm \Theta_{\ell-1}, \one\{P \bm \Theta_{\ell-1} > c_r\})) \in A, \\\
 & \qquad \qquad
               P \max\{1, r(\bm \Theta_1), \ldots, r(\bm \Theta_{\ell-1})\} \leq c_r, \,
               P \min\{r(\bm \Theta_{-1}), r(\bm \Theta_\ell)\} > c_r\big). 
 \end{align*}	
Analogously to former calculations, the first term equals 
$$c_r^{-\alpha} \EE[\min\{1, r(\bm \Theta_{1})^\alpha, \ldots, r(\bm \Theta_{\ell-1})^\alpha\} 
- \max\{r(\bm \Theta_{-1})^\alpha,r(\bm \Theta_\ell)^\alpha\}]_+.$$
Substituting $\eta = (c_r/y)^\alpha$ and using similar arguments as in the proof of Proposition \ref{prop:spat-risk}, the second term can be rewritten to
\begin{align*}
& c_r^{-\alpha} \PP\big( (\tilde c(\bm \Theta_0, \{\bm \Theta_0^\alpha > \eta\}), \ldots, \tilde c(\bm \Theta_{\ell-1}, \{\bm \Theta_{\ell-1}^\alpha > \eta\})) \in A,\\
& \qquad \qquad \qquad r(\bm \Theta_{-1})^\alpha \leq \eta, r(\bm \Theta_0)^\alpha > \eta, \ldots, r(\bm \Theta_{\ell-1})^\alpha > \eta, r(\bm \Theta_\ell)^\alpha \leq \eta \big)
\, \mathrm{d}\eta.
\end{align*}
\qed

\end{document}